\documentclass[superscriptaddress,preprint,eqsecnum,
amssymb,amsmath,nobibnotes,aps,prd,showkeys,showpacs,nofootinbib]{revtex4}

\usepackage[latin1]{inputenc}
\usepackage{graphicx}
\usepackage[english]{babel}

\usepackage{amsmath}
\usepackage{amssymb}
\usepackage{amsfonts}
\usepackage{colordvi}
\usepackage{psfrag}
\usepackage{color}

\begin{document}

\def\cL{{\cal L}}
\def\be{\begin{equation}}
\def\ee{\end{equation}}
\def\bea{\begin{eqnarray}}
\def\eea{\end{eqnarray}}
\def\beq{\begin{eqnarray}}
\def\eeq{\end{eqnarray}}
\def\tr{{\rm tr}\, }
\def\nn{\nonumber \\}
\def\e{{\rm e}}
\def\bef{\begin{figure}}
\def\eef{\end{figure}}
\newcommand{\ans}{ansatz }
\newcommand{\eeqn}{\end{eqnarray}}
\newcommand{\bd}{\begin{displaymath}}
\newcommand{\ed}{\end{displaymath}}
\newcommand{\mat}[4]{\left(\begin{array}{cc}{#1}&{#2}\\{#3}&{#4}
\end{array}\right)}
\newcommand{\matr}[9]{\left(\begin{array}{ccc}{#1}&{#2}&{#3}\\
{#4}&{#5}&{#6}\\{#7}&{#8}&{#9}\end{array}\right)}
\newcommand{\matrr}[6]{\left(\begin{array}{cc}{#1}&{#2}\\
{#3}&{#4}\\{#5}&{#6}\end{array}\right)}
\newcommand{\cvb}[3]{#1^{#2}_{#3}}
\def\lsim{\raise0.3ex\hbox{$\;<$\kern-0.75em\raise-1.1ex
e\hbox{$\sim\;$}}}
\def\gsim{\raise0.3ex\hbox{$\;>$\kern-0.75em\raise-1.1ex
\hbox{$\sim\;$}}}
\def\abs#1{\left| #1\right|}
\def\simlt{\mathrel{\lower2.5pt\vbox{\lineskip=0pt\baselineskip=0pt
           \hbox{$<$}\hbox{$\sim$}}}}
\def\simgt{\mathrel{\lower2.5pt\vbox{\lineskip=0pt\baselineskip=0pt
           \hbox{$>$}\hbox{$\sim$}}}}
\def\unity{{\hbox{1\kern-.8mm l}}}
\newcommand{\eps}{\varepsilon}
\def\ep{\epsilon}
\def\ga{\gamma}
\def\Ga{\Gamma}
\def\om{\omega}
\def\omp{{\omega^\prime}}
\def\Om{\Omega}
\def\la{\lambda}
\def\La{\Lambda}
\def\al{\alpha}
\newcommand{\ov}{\overline}
\renewcommand{\to}{\rightarrow}
\renewcommand{\vec}[1]{\mathbf{#1}}
\newcommand{\vect}[1]{\mbox{\boldmath$#1$}}
\def\tm{{\widetilde{m}}}
\def\mcirc{{\stackrel{o}{m}}}
\newcommand{\Dm}{\Delta m}
\newcommand{\dm}{\varepsilon}
\newcommand{\tanb}{\tan\beta}
\newcommand{\nbar}{\tilde{n}}
\newcommand\PM[1]{\begin{pmatrix}#1\end{pmatrix}}
\newcommand{\up}{\uparrow}
\newcommand{\down}{\downarrow}
\def\omE{\omega_{\rm Ter}}

\newcommand{\Dsusy}{{susy \hspace{-9.4pt} \slash}\;}
\newcommand{\DCP}{{CP \hspace{-7.4pt} \slash}\;}
\newcommand{\mc}{\mathcal}
\newcommand{\gr}{\mathbf}
\renewcommand{\to}{\rightarrow}
\newcommand{\gtc}{\mathfrak}
\newcommand{\wh}{\widehat}
\newcommand{\br}{\langle}
\newcommand{\kt}{\rangle}

\def\lsim{\mathrel{\mathop  {\hbox{\lower0.5ex\hbox{$\sim$}
\kern-0.8em\lower-0.7ex\hbox{$<$}}}}}
\def\gsim{\mathrel{\mathop  {\hbox{\lower0.5ex\hbox{$\sim$}
\kern-0.8em\lower-0.7ex\hbox{$>$}}}}}

\def\nn{\\  \nonumber}
\def\de{\partial}
\def\brf{{\mathbf f}}
\def\bbf{\bar{\bf f}}
\def\bF{{\bf F}}
\def\bbF{\bar{\bf F}}
\def\bA{{\mathbf A}}
\def\bB{{\mathbf B}}
\def\bG{{\mathbf G}}
\def\bI{{\mathbf I}}
\def\bM{{\mathbf M}}
\def\bY{{\mathbf Y}}
\def\bX{{\mathbf X}}
\def\bS{{\mathbf S}}
\def\bb{{\mathbf b}}
\def\bh{{\mathbf h}}
\def\bg{{\mathbf g}}
\def\bla{{\mathbf \la}}
\def\bmu{\mathbf m }
\def\by{{\mathbf y}}
\def\bmu{\mbox{\boldmath $\mu$} }
\def\bsig{\mbox{\boldmath $\sigma$} }
\def\bunity{{\mathbf 1}}
\def\cA{{\cal A}}
\def\cB{{\cal B}}
\def\cC{{\cal C}}
\def\cD{{\cal D}}
\def\cF{{\cal F}}
\def\cG{{\cal G}}
\def\cH{{\cal H}}
\def\cI{{\cal I}}
\def\cL{{\cal L}}
\def\cN{{\cal N}}
\def\cM{{\cal M}}
\def\cO{{\cal O}}
\def\cR{{\cal R}}
\def\cS{{\cal S}}
\def\cT{{\cal T}}
\def\eV{{\rm eV}}

\title{Nonlocal quantum field theory without acausality and nonunitarity at quantum level:
is SUSY the key?}

\author{Andrea Addazi}

\affiliation{Dipartimento di Fisica,
Universit\`a di L'Aquila, 67010 Coppito AQ, Italy}
 
\affiliation{Laboratori Nazionali del Gran Sasso (INFN), 67010 Assergi AQ, Italy}
 
\author {Giampiero Esposito}
 
\affiliation{Istituto Nazionale di Fisica Nucleare,  
Sezione di Napoli, Complesso Universitario di
Monte S. Angelo, Via Cintia Edificio 6, 80126, Napoli, Italy}

\date{\today}

\begin{abstract}
The realization of a nonlocal quantum field theory 
without losing unitarity, gauge invariance and causality is investigated.
It is commonly retained that such a formulation is possible at tree level, 
but at quantum level acausality is expected to reappear at one loop. 
We suggest the the problem of acausality is, 
in a broad sense, similar to the one about anomalies in quantum field theory. 
By virtue of this analogy, we suggest that acausal diagrams
resulting from the fermionic sector and the bosonic one
might cancel each other, with a suitable 
content of fields and suitable symmetries. 
As a simple example, we show how supersymmetry can alleviate this problem
in a simple and elegant way, {\it i.e.}, by leading to exact
cancellations of harmful diagrams, to all orders of perturbation theory.
An infinite number of divergent diagrams cancel each other by virtue 
of the nonrenormalization theorem of supersymmetry. 
However, supersymmetry is not enough to protect 
a theory from all acausal divergences.
For instance, acausal contributions to supersymmetric 
corrections to $D$-terms are not protected by supersymmetry.
On the other hand, we show in detail how supersymmetry also helps 
in dealing with $D$-terms: divergences are not cancelled but they become 
softer than in the nonsupersymmetric case. The supergraphs' formalism turns out
to be a powerful tool to reduce the complexity of perturbative calculations.
\end{abstract}

\maketitle

\section{Introduction}

The successes of local quantum field theories are well known. 
They make it possible to understand the current results in particle physics.
The first famous success of a local quantum field theory
was the capability to account for the anomaly in the magnetic moment
of the electron, with respect to the prediction of Dirac's equation,
calculated independently by Tomonaga \cite{Tomonaga}, Schwinger \cite{Schwinger1,Schwinger2} 
and Feynman \cite{Feynman1,Feynman2,Feynman3}.
However, one-loop calculations like these lead to divergences.
These are originating from locality: 
we are assuming, in a local quantum field theory, that quantum fields are interacting 
at a vertex point in spacetime.
On the other hand, we can control these divergences through renormalization techniques,
with the subtraction of counterterms. Among physicists,
the most famous skeptical about this practice was Dirac himself \cite{Dirac}.
He viewed renormalization as a trick without mathematical consistency.
On the other hand, renormalization has a well understood physical meaning 
in the interpretation of quantum field theories as effective field theory,
formalized through the renormalization group approach.
In fact the infinities are cutoff by a scale $\Lambda$, 
possibly coincident with the Planck scale of quantum gravity or so,
where an unknown new physics, supposed completely different 
from local quantum field theories, has to be considered. 
This was inspired by condensed matter physics.
In fact, in a material, it is straightforward to understand why a 
a physical short-distance regulator has to be considered,
because of the transition from the continuum to the atomic discretization. 
In this sense, the Dirac doubts were premature for his times. 
However, the various different approaches to quantum gravity
like string theory \cite{GSW}, loop quantum gravity \cite{loop}, 
noncommutative geometry \cite{noncommutative}
seem to lead again to the same problem: nonlocality in a quantum field theory.
In fact, string \cite{higher} theory\footnote{For string-inspired phenomenology see Ref. 
\cite{Addazi:2014ila,Addazi:2015ata,Addazi:2015rwa,Addazi:2015hka}. Exotic stringy 
instantons can indeed induce nonperturbative couplings 
leading to new physical effects, for example in ultra-cold neutron physics. 
These effects are computable and perfectly controllable in a large class of string models, leading to effects that 
cannot be generated at all in gauge theories.} 
is intrinsically nonlocal, and the presence of Kaluza-Klein towers of infinitely 
many higher-spin fields is a general feature, leading to 
nonlocalities\footnote{Nonlocal classical infrared extensions of General Relativity
were also proposed in order to simulate, dynamically, the apparent presence of dark energy in our 
Universe \cite{ArkaniHamed:2002fu}. See also \cite{NL1,NL2} for recent developments in this subject. 
Similarly, Lorentz-violating infrared extensions of the Einstein-Hilbert action might be another viable 
alternative to nonlocal gravity. For instance, a Lorentz violating mass term for 
the graviton can be added to the Einstein-Hilbert action without the introduction of ghosts 
\cite{LIV1,LIV2,LIV3}. However, other phenomenological difficulties, 
such as geodetic instabilities around stars and black holes for large part of the parameters' space, 
reject a large class of these extensions \cite{Addazi:2014mga}. }. For example, 
we remember the case of a spin $s=3$ field, with nonlocal equation of motion:
a local quantum field theory approach does not permit to quantize such a theory. 
On the other hand, in loop quantum gravity \cite{loop}, the ``discretization" of spacetime
itself naturally leads to the question about the presence of quantum fields of matter
in such a spacetime. In noncommutative geometry \cite{noncommutative}, 
the peculiar topology is inducing new 
nonlocal interaction terms in the functional integral of matter fields. 
This could suggest that the next step of the road to quantum gravity 
might be to solve the question about the admissibility of a nonlocal quantum field theory.  
A lot of attempts to formulate a quantum field theory without locality 
have been made, in order to evade the problem of ultraviolet divergencies 
in a consistent way. The idea is to circumvent the ultraviolet 
infinities by replacing local vertices with nonlocal, smeared ones. 
The first suggestion was made by Wataghin in Ref. \cite{Wataghin}.
However, it was realized that a lot of problems
in a nonlocal quantum field theory have to be solved
in order to have a meaningful and computable theory
(for example, see the review in Ref. \cite{review}). 
The main difficulty is to construct a theory preserving
unitarity, gauge invariance and causality at the same time
\cite{Efimov1,Moffat,W1,W2,W3,Evens}.  
Unitarity of the S-matrix is a necessary condition
for a ``meaningful theory" that evaluates probabilities. 
For example, this condition automatically saves the theory from negative-norm states on-shell. 
On the other hand, nonlocal interactions have to be compatible with a local gauge symmetry:
nonlocalization of the vertices has to be compatible 
with the essential consequences of a local gauge symmetry. 
Eventually, causality just requires that the commutator between 
two fields, located at two different spacetime positions,
has to vanish for a spacelike distance between the two spacetime locations. 
A consistent formulation of a nonlocal quantum field theory
preserving unitarity, local gauge symmetry and causality 
at tree level was presented in Refs. \cite{W1,W2,Moffat,Evens,W3} as
a $\lambda \phi^{4}$ model, simply generalizable
to a Yang-Mills model in the same paper just cited above. 
These models are realized through the introduction
of auxiliary fields, not existing on-shell but to be considered 
off-shell (at least with exception of few particular cases of gauge choices).
Unfortunately, as shown in Refs. \cite{Joglekar1,Joglekar2},
such models are losing causality just at one loop. 
In particular, in the simplest case of a massless self-interacting model $\lambda\phi^{4}$, 
one-loop diagrams with propagating auxiliary fields
lead to divergences: in a scattering process $\phi\phi\rightarrow \phi\phi$, 
the divergence occurs  
at lower order $\mathcal{S}\sim s^{2}+t^{2}+u^{2}$, where $\mathcal{S}$ is the $S$-matrix 
describing such a scattering. 

Is it possible to avoid this causality violation at quantum level? 
This is the issue that we are considering in this paper.
In fact, we would like to suggest a possible analogy with the problem of the
{\it anomalies} in quantum field theory. 
Anomalies are in general the loss of a gauge symmetry 
of a classical action, when it is considered at quantum level. 
For instance, in the standard model, we have an 
anomalous contribution from triangle diagrams, 
but all contributions resulting from quarks and leptons 
cancel each other, leading to a consistent theory. 
We would like to stress that the electroweak theory
$SU(2)\times U(1)$ proposed by Glashow, Weinberg, Salam
would be inconsistent because of local anomalies if we were not considering the 
counter contributions from the quark sector
- in this sense, the electroweak theory naturally suggests
its own gauge-group extension. 
Thus, our attitude about acausalities could be the same: 
we have to construct a nonlocal quantum field theory
with a field content leading to automatic cancellations
of harmful quantum diagrams. 

In this paper, we are suggesting that {\it a supersymmetric nonlocal quantum field theory 
is automatically free of causality violations resulting from supersymmetric $F$-terms to all orders
of perturbation theory}. This is a direct consequence 
of the {\it nonrenormalization theorem} of supersymmetry \cite{NR1,NR2}.
If F-terms cannot be corrected by radiative contributions, 
automatically we will obtain a cancellation of all acausal $n$-loop contributions.
In particular, we consider a susy generalization of 
the Eliezer-Woodard-Moffat-Kleppe-Evans
model, described in Refs. \cite{W1,W2,Moffat,Evens,W3} for a $\lambda \phi^4$ model
and a gauge theory. We also stress that our result is valid only under 
the particular prescription described in Refs.
\cite{W1,W2,Moffat,Evens,W3} \footnote{An alternative approach 
to the one considered in this paper was 
studied in \cite{Anumap1,Anumap2,Anumap3,B1}, with several implications in cosmology and LHC. 
In this case, issues about gauge invariance and unitarity are not 
considered as a problem: an effective string-inspired approach is developed. 
Recently, questions about string theory 
and causality were considered in Refs. \cite{Maldacena} and
\cite{Veneziano}. In particular, the scattering 
of gravitons in Regge limit was considered
and the entire Kaluza-Klein tower appears to 
cure acausal contributions \cite{Veneziano}.
Questions about causality at all orders of perturbations 
in string theory remain open. 
For useful discussions about String amplitudes and 
possible LHC implications see also Refs. \cite{Bianchi1,Bianchi2}.}.
Curiously, a supersymmetric generalization of EWMKE was not considered 
before in literature. We would like to clarify that
our model cannot be considered in a Lorentz violating space-time and non-commutative geometries.
In particular, we will consider a nonlocal Wess-Zumino
and a nonlocal super Yang-Mills, showing how $F$-terms remain causal forever. 
Unfortunately, acausalities will reappear by means of $D$-terms
in a nonlocal super Yang-Mills: the nonrenormalization theorem does not protect 
the theory against these corrections. 
For this reason, we suggest that only an infinite number of bosons 
and fermions can cancel all the infinite number of dangerous divergences.
In this sense, supersymmetry is a powerful framework in order to cancel 
an infinite number of harmful diagrams, but 
unfortunately it is not enough to cancel all the acausalities. 
However, we show that supersymmetry also gets softer 
acausal violations with respect to the nonsupersymmetric case. 
For these reasons, supersymmetry seems to be 
the key for the realization of a nonlocal quantum field theory.

In Sec. II, we build the nonlocal Wess-Zumino model.
In Sec. III, we discuss the main points about the consistency 
of this model, considering gauge transformations, unitarity, causality.
In Secs. IV-V-VI, we get the main result of the paper: 
the cancellations of harmful acausal loops resulting from $F$-terms. 
In Sec. VII, we discuss the generalizations of our results about Wess-Zumino models
by considering a nonlocal super Yang-Mills model, discussing also how in this case the
cancellations occur for F-terms. In Sec. VIII, we are ending with our comments and perspectives. 
Thus, for the self-consistency of our paper, we find it necessary 
to summarize the basic properties of local Wess-Zumino \cite{WB} 
and local super Yang-Mills models, nonlocal nonsupersymmetric $\lambda \phi^{4}$ model
in Appendices, in order to facilitate the logical step towards their nonlocal generalization. 

\section{Nonlocal Wess-Zumino model}

In this section, we are generalizing the local Wess-Zumino model ($\mathcal{N}=1$ susy) in (\ref{kin})-(\ref{int}), 
to a nonlocal one. 
We are introducing an extra chiral supermultiplet 
$\Phi^{2}=\{F^{2},\phi^{2},\psi^{2} \}$ playing the role of auxiliary superfield, 
in order to construct a nonlocal model, without losing rigid supersymmetry. 
In other words, we are extending a nonlocal scalar model invariant under the global Poincar\'e 
group to a nonlocal chiral superfield model invariant under the global superPoincar\'e group. 

We consider the following auxiliary supersymmetric action\footnote{We are assuming for the 
moment a number of spacetime dimensions $D=4$, but this formalism can be 
extended to an arbitrary number of spacetime dimensions. This framework 
might be of interest for string theories.}:
\begin{equation} 
\label{SWZ}
S[\hat{F}^{1},F^{2},\hat{\phi}^{1}, \phi^{2},\hat{\psi}^{1},\psi^{2}]
=\mathcal{F}[\hat{F}^{1},\hat{\phi}^{1},\hat{\psi}^{1}]
-\mathcal{A}[F^{2},\phi^{2},\psi^{2}]
+\mathcal{I}[{\hat F}^{1}+F^{2},{\hat \phi}^{1}+\phi^{2},{\hat \psi}^{1}+\psi^{2}],
\end{equation}
with 
\begin{equation} 
\label{Fphi}
\mathcal{F}[\hat{F}^{1},\hat{\phi}^{1},\hat{\psi}^{1}]
=\int {\rm d}^{4}x [\hat{F}^{1^{\dagger}}_{i}a_{ij}^{1}\hat{F}^{1}_{j}
+\hat{\phi}^{1^{\dagger}}_{i}b^{1}_{ij}\hat{\phi}^{1}_{j}
+\bar{\hat{\psi}}^{1^{\dagger}}_{i}f^{1}_{ij}\hat{\psi}^{1}_{j}],
\end{equation}
\begin{equation} 
\label{Aphi}
\mathcal{A}[F^{2},\phi^{2},\psi^{2}]=\int {\rm d}^{4}x [F^{2^{\dagger}}_{i}a_{ij}^{2}F^{2}_{j}
+\phi^{2^{\dagger}}_{i}(x)b^{2}_{ij}\phi^{2}_{j}(x)
+\bar{\psi}^{2^{\dagger}}_{i}f^{2}_{ij}\psi_{j}^{2}],
\end{equation}
where $\hat{\phi}^{1},\hat{\psi}^{1}, \hat{F}^{1}$ are the ``smeared fields".
For simplicity, we might choose an exponential smearing operator for the scalar field defined as 
\begin{equation} 
\label{phiat}
\hat{\phi}_{1} \equiv \mathcal{E}^{-1}\phi_{1},\,\,\,\,\,\mathcal{E} \equiv {\rm e}^{b^{1}/2\Lambda^{2}}.
\end{equation}
Automatically, in order to preserve the (rigid) supersymmetry transformations
\begin{equation}
\label{susy1}
\delta_{\zeta} \phi_{i}^{1,2}=\zeta \psi_{i}^{1,2},
\end{equation}
\begin{equation}
\label{susy2}
\delta_{\zeta} \psi_{i}^{1,2}=-{\rm i}\sigma^{\mu}{\rm i}\sigma_{2}\zeta^{*}\partial_{\mu}\phi_{i}^{1,2}
+\zeta F_{i}^{1,2},
\end{equation}
\begin{equation}
\label{susy3}
\delta_{\zeta} F_{i}^{1,2}=-{\rm i}\zeta^{*}\bar{\sigma}^{\mu}\partial_{\mu}\psi_{i}^{1,2},
\end{equation}
we have to take rigidly the same smearing prescription for the superpartners, {\it i.e.}
\begin{equation} 
\label{psiat}
\hat{\psi}^{1} \equiv \mathcal{E}^{-1}\psi_{1},\,\,\,\hat{F}^{1} \equiv \mathcal{E}^{-1}F^{1}.
\end{equation}

In (\ref{SWZ}) we also define the following prescriptions for auxiliary fields ($I$ being
the identity operator) 
\begin{equation} 
\label{a1}
a^{2} \equiv \frac{a^{1}}{(\mathcal{E}^{2}-I)},\,\,\,\,
b^{2} \equiv \frac{b^{1}}{(\mathcal{E}^{2}-I)},\,\,\,
f^{2} \equiv \frac{f^{1}}{(\mathcal{E}^{2}-I)}.
\end{equation}
The action of the nonlocal theory is obtained as
\begin{equation} 
\label{Sphipsi}
\hat{S}[\hat{F}^{1},\hat{\phi}^{1},\hat{\psi}^{1}] 
=S\left[\hat{F}^{1},\hat{\phi}^{1},\hat{\psi}^{1},F^{2}[\hat{F}^{1},\hat{\phi}^{1},\hat{\psi}^{1}],
\phi^{2}[\hat{F}^{1},\hat{\phi}^{1},\hat{\psi}^{1}],
\psi^{2}[\hat{F}^{1},\hat{\phi}^{1},\hat{\psi}^{1}]\right],     
\end{equation}
where $F^{2}[\hat{F}^{1},\hat{\phi}^{1},\hat{\psi}^{1}]$, $\phi^{2}[\hat{F}^{1},\hat{\phi}^{1},\hat{\psi}^{1}]$,
$\psi^{2}[\hat{F}^{1},\hat{\phi}^{1},\hat{\psi}^{1}]$ solve the classical equations
\begin{equation} 
\label{Spsiphi}
\frac{\delta {S}}{\delta \phi^{2}}=\frac{\delta {S}}{\delta \psi^{2}}=\frac{\delta {S}}{\delta F^{2}}=0.
\end{equation}

Of course, we can also rewrite the auxiliary action (\ref{SWZ}) in a form explicitly invariant under (rigid) supersymmetry:
\begin{equation} 
\label{expsusy}
\int {\rm d}^{4}x \Bigr[ \mathcal{G}\Bigr\{\int {\rm d}^{2}\theta 
{\rm d}^{2}\bar{\theta}\Bigr(\mathcal{K}^{1}(\bar{\Phi}^{1},\Phi^{1})
+\mathcal{K}^{2}(\bar{\Phi}^{2},\Phi^{2})\Bigr)\Bigr\}
+\int {\rm d}^{2}\theta \mathcal{W}(\Phi^{1}+\Phi^{2})
+\int {\rm d}^{2}\bar{\theta} \bar{\mathcal{W}}(\bar{\Phi}^{1}+\bar{\Phi}^{2})\Bigr],
\end{equation}
where $\mathcal{K}^{1,2}$ and $\mathcal{G}\{\mathcal{O} \}$ 
are encoding the nonlocal bilinear terms, 
for the standard and the auxiliary chiral superfields\footnote{
Naturally, the prescription for $\mathcal{K}^{2}$ 
encodes the extra sign in the nonlocal kinetic terms for the auxiliary fields, 
$\mathcal{A}$ in (\ref{SWZ}), with respect to the standard fields' term $\mathcal{F}$.}.
Let us recall that 
\begin{equation}
\label{remind1}
\Phi=\phi+\sqrt{2}\theta \psi(x)+{\rm i}\theta \sigma^{\mu}\bar{\theta}\partial_{\mu}\phi(x)
-\theta \theta F(x)-\frac{{\rm i}}{\sqrt{2}}\theta \theta \partial_{\mu}\psi(x) \sigma^{\mu}\bar{\theta}
-\frac{1}{4}\theta\theta \bar{\theta}\bar{\theta}\Box \phi(x),
\end{equation}
and 
\begin{equation}
\label{remind2}
\bar{\Phi}=\bar{\phi}+\sqrt{2}\bar{\theta} \bar{\psi}(x)+{\rm i}\bar{\theta} \sigma^{\mu}
\theta\partial_{\mu}\bar{\phi}(x)-\bar{\theta} \bar{\theta} F(x)
-\frac{{\rm i}}{\sqrt{2}}\bar{\theta} \bar{\theta} \partial_{\mu}\bar{\psi}(x) \sigma^{\mu}\theta
-\frac{1}{4}\theta\theta \bar{\theta}\bar{\theta}\Box \bar{\phi}(x),
\end{equation}
so that 
\begin{equation}
\label{remind3}
\int {\rm d}^{2}\theta {\rm d}^{2}\bar{\theta}\bar{\Phi}\Phi=\partial_{\mu}\phi^{*}\partial^{\mu}\phi
+{\rm i}\bar{\psi}\sigma^{\mu}\partial_{\mu}\psi+FF^{*},
\end{equation}
and essentially $\mathcal{G},\mathcal{K}_{1,2}$ map (\ref{remind3}) for $\Phi^{(1,2)}$ into 
\begin{eqnarray}
\label{map1}
\; & \; & 
\left(1-\sum_{n}\frac{1}{n!}(\partial^{\mu} \phi^{(1)^{*}}\partial_{\mu} \phi^{(1)})^{n}\right)
+\left(1-\sum_{n}\frac{1}{n!}({\rm i}\bar{\psi}^{(1)}\sigma^{\mu}\partial_{\mu}\psi^{(1)})^{n}\right)
\nonumber \\
&+& \Box^{-1}\left(1-\sum_{n}\frac{1}{n!}(F^{(1)^{*}}\Box F^{(1)})^{n}\right)
\nonumber \\
&+& \sum_{n,k}\frac{1}{k!}(n\Box)^{k}(\partial^{\mu} \phi^{(2)^{*}}\partial_{\mu} \phi^{(2)})
+\sum_{n,k}\frac{1}{k!}(n\Box)^{k}({\rm i}\bar{\psi}^{(2)}\sigma^{\mu}\partial_{\mu}\psi^{(2)})
\nonumber \\
&+& \sum_{n,k}\frac{1}{k!}(n\Box)^{k}(F^{(2)^{*}} F^{(2)}).
\end{eqnarray}
Thus, as an example, a suitable set of functions in (\ref{expsusy}) is as follows:
\begin{equation}
\label{suitable1}
\mathcal{K}^{1}(t)=\mathcal{K}^{2}(t)=t,
\end{equation}
with $t$ generic variable of these functions, {\it i.e.}, they
correspond to minimal K\"{a}hler potentials
$\Phi^{(1,2)^{\dagger}}\Phi^{(1,2)}$. Moreover, one has 
\begin{eqnarray}
\label{suitable2}
\mathcal{G}(\mathcal{X}^{(1)},\mathcal{Y}^{(1)},\mathcal{Z}^{(1)},\mathcal{X}^{(2)},\mathcal{Y}^{(2)},\mathcal{Z}^{(2)})
&=&
({\rm e}^{\mathcal{X}^{(1)}}-1)+({\rm e}^{\mathcal{Y}^{(1)}}-1)+\Box^{-1}({\rm e}^{\mathcal{Z}^{(1)}}-1)
\nonumber \\
&+& \mathcal{F}_{1}(\mathcal{X}^{(2)})
+\mathcal{F}_{2}(\mathcal{Y}^{(2)})
+\mathcal{F}_{3}(\mathcal{Z}^{(2)}),
\end{eqnarray}
where $\mathcal{F}_{1,2,3}$ are obtained from the following power series:
\begin{equation}
\label{F1F2F3}
\mathcal{F}_{1,2,3}(\mathcal{X}^{(2)},\mathcal{Y}^{(2)},\mathcal{Z}^{(2)})
=\sum_{k,n=0}^{\infty}\frac{n^{k}}{k!}\Box^{k}(\mathcal{X}^{(2)},\mathcal{Y}^{(2)},\Box^{-1}\mathcal{Z}^{(2)}),
\end{equation}
while 
\begin{equation}
\label{X1}
\mathcal{X}^{(1,2)} \equiv \partial^{\mu} \phi^{(1,2)^{*}}\partial_{\mu} \phi^{(1,2)},
\end{equation}
\begin{equation}
\label{Y1}
\mathcal{Y}^{(1,2)} \equiv {\rm i}\bar{\psi}^{(1,2)}\sigma^{\mu}\partial_{\mu}\psi^{(1,2)},
\end{equation}
\begin{equation}
\label{Z1}
\mathcal{Z}^{(1,2)} \equiv F^{(1,2)^{*}} \Box F^{(1,2)}.
\end{equation}

\subsection{Feynman rules}

In such a model, simple Feynman rules can be consistently 
defined, as a trivial extension of what holds for ordinary Feynman diagrams.
For more details about important technical aspects in the quantization procedure 
see Appendix E, and for readers who do not feel satisfied as yet, we recall that Refs. 
\cite{W1,W2,Moffat,Evens,W3} have considered these features in nonlocal (nonsupersymmetric)
gauge theories.

The vertices are unchanged, and the smeared propagator for the fields $\phi^{1},\psi^{1}$ reads as 
\begin{equation} 
\label{pro1}
\frac{{\rm i}\mathcal{E}^{2}}{(b^{1}+{\rm i}\epsilon)},\,\,\,\frac{{\rm i}\mathcal{E}^{2}}{(f^{1}+{\rm i}\epsilon)}.
\end{equation}
Moreover, the smeared propagators for the auxiliary\footnote{Also in this case, as usual in 
quantum field theory, the Feynman rules for Majorana fermions are reduced to 
Dirac fermion ones, because of the cancellations between the charge conjugation 
$C$ operators occurring in the Majorana propagator and in the vertices involving it. 
For this reason, we omit the presence of the $C$ operators in the given rules.}
fields $\phi^{2},\psi^{2}$ are
\begin{equation} 
\label{pro2}
\frac{{\rm i}[I-\mathcal{E}^{2}]}{(b^{1}+{\rm i}\epsilon)},\,\,\,
\frac{{\rm i}[I-\mathcal{E}^{2}]}{(f^{1}+{\rm i}\epsilon)}.
\end{equation}

On considering the familiar form of the differential operators
\begin{equation} 
\label{b1}
b_{1}=\partial^{2}-m^{2},\,\,\,\,\,f_{1}={\rm i}\gamma^{\mu}\partial_{\mu}+m,
\end{equation}
the corresponding rules in momentum space are as follows.

i) For the $\phi^{1},\psi^{1}$ propagator: 
\begin{equation} 
\label{prop1psi1}
{\rm i}\frac{ {\rm exp}\left( \frac{-p^{2}-m^{2}}{\Lambda^{2}}\right) } {(p^{2}+m^{2}+{\rm i}\epsilon)},\,\,\,\,
{\rm i}\frac{ {\rm exp}\left( \frac{-p^{2}-m^{2}}{\Lambda^{2}}\right) }  
{(\gamma_{\mu}p^{\mu}-m+{\rm i}\epsilon)},
\end{equation}

ii) For the $\phi^{2},\psi^{2}$ propagator: 
\begin{equation} 
\label{prop2psi2}
{\rm i}\frac{ \left[I-{\rm exp}\left( \frac{-p^{2}-m^{2}}{\Lambda^{2}}\right) \right]}
{(p^{2}+m^{2}+{\rm i}\epsilon)}, \,\,\,\,{\rm i}\frac{ \left[I-{\rm exp}\left( 
\frac{-p^{2}-m^{2}}{\Lambda^{2}}\right) \right]}{(\gamma_{\mu}p^{\mu}-m+{\rm i}\epsilon)}.
\end{equation}
Equations (\ref{prop1psi1}) and (\ref{prop2psi2}) provide the integrand in the integral representation of 
the propagator \footnote{We would like to note that
one can also redefine all propagators with an exchange sign in the exponent 
of $\mathcal{E}$: as we will see, this corresponds to 
introduce polinomial divergences 
through ordinary fields $\phi^{(1)},\psi^{(1)}$
rather  than through $\phi^{(2)},\psi^{(2)}$.}. 
We note, incidentally, that such an integrand is the inverse of the symbol
$$
{(p^{2}+m^{2})\over (I-{\rm e}^{(-p^{2}-m^{2})\over \Lambda^{2}})}.
$$
We are therefore dealing with pseudodifferential
operators, a topic well known in the mathematical literature \cite{Widom,Gilkey,Booss}.

\section{Consistency of the model}

In this section we discuss the consistency of our model.
In particular, we construct a nonlocal quantum field theory 
which is unitary to all orders of perturbation 
theory and is invariant under local symmetry transformations.
Indeed, in general a nonlocal theory respects causality and unitarity only at tree level.
However, we will show how supersymmetry automatically 
guarantees causality and unitarity to all orders of perturbation 
theory\footnote{Details about the consistency of functional-integral quantization
in a nonlocal field theory are discussed in Ref. \cite{W3}, especially the problem 
of the measure in such functional integrals.}.

\subsection{Symmetry transformations}

Generically, a nonlocal quantum field theory seems 
hopeless: how to construct a local gauge symmetry?
However, gauge symmetry can be encoded in 
a nonlocal theory with a new nonlinear transformation 
rule. In fact, as shown in Ref. \cite{W3} for the scalar theory, 
if an infinitesimal transformation
$\delta \phi^{1}_{i}=T_{i}[\phi^{1}]$
generates a symmetry of the local action $S[\phi^{1}]$, then a transformation
$\hat{\delta}\phi^{1}_{i}=\mathcal{E}_{ij}^{2}T_{j}[\phi^{1}+\phi^{2}[\phi^{1}]]$
generates a symmetry for the corresponding nonlocal action $\hat{S}[\phi]$. 
In a broad sense, the procedure for obtaining a nonlocal theory preserves 
a deformed version of the usual continuous symmetry, and we can write 
\begin{equation} 
\label{auxialiar}
\hat{\delta}\phi_{i}^{2}[\phi^{1}]=\left(I-\mathcal{E}^{2}\right)_{ij}T_{j}\left[\phi^{1}
+\phi^{2}[\phi^{1}]\right]-K_{ij}\left[\phi^{1}+\phi^{2}[\phi^{1}]\right]\frac{\delta T_{k}}
{\delta \phi_{j}^{1}}\left[\phi^{1}+\phi^{2}[\phi^{1}] \right]\mathcal{E}_{kl}^{2}
\frac{\delta \hat{S}[\phi^{1}]}{\delta \phi^{1}_{l}},
\end{equation}
\begin{equation} 
\label{Ktr}
K_{ij}^{-1}[\phi^{1}]=b^{2}_{ij}-\frac{\delta^{2}\mathcal{I}[\phi]}{\delta \phi_{i}\delta \phi_{j}}.
\end{equation}
For our supersymmetric model, this property remains unaffected, 
in a well understood way, exactly as it occurs in the usual supersymmetric gauge models.

\subsection{Unitarity and causality}

Unitarity is necessary for the consistency and the calculability 
of a model. This means that the total probability of processes studied in quantum field theory
has to be conserved and equal to one. In other words, the $S$-matrix has to satisfy the condition 
\begin{equation}
\label{Smatrix}
{\mathcal S}{\mathcal S}^{\dagger}=
{\mathcal S}^{\dagger}{\mathcal S}=I.
\end{equation}
On the other hand, causality is fundamental for an obvious reason:
without it, a quantum field theory model could never make any predictions
testable against observation.
Imagine a theory without causality: in order to calculate the probability of a collision 
occurring, for example, at LHC, we would have to know the status
of all the rest of the Universe!
In other words, causality is strictly connected to the {\it cluster decomposition principle} 
of the S-matrix \cite{Weinberg}:
if multi-particle processes $a_{1}\rightarrow b_{1},\,\,a_{2}\rightarrow b_{2},...,\,\,a_{N}\rightarrow b_{N}$ 
are studied in $N$ laboratories at spacetime positions $x_{1,..,N}$ causally disconnected,
for which therefore $(x_{i}-x_{j})^{2}<0$ ($i\neq j$, $i,j=1,...,N$), 
then the S-matrix is factorized into $N$ parts according to 
\begin{equation} 
\label{Cluster}
\mathcal{S}_{b_{1}+b_{2}+...b_{N},a_{1}+a_{2}+...+a_{N}}
=\mathcal{S}_{b_{1}a_{1}}\mathcal{S}_{b_{2}a_{2}}...\mathcal{S}_{b_{N}a_{N}}.
\end{equation}
From (\ref{Cluster}), we are able to say that quantum fields 
are commuting at spacelike distances: the causality violation condition is expressed as
\begin{equation} 
\label{phiphi}
[\phi(x),\phi(y)]=0,\,\,\,\,\,(x-y)^{2}<0,
\end{equation}
which is re-expressed as a microcausality condition \cite{BS} on the S-matrix as
\begin{equation} 
\label{transl}
\frac{\delta}{\delta \phi(x)}\left(\frac{\delta {\mathcal S}[\phi]}{\delta \phi(y)}
{\mathcal S}^{\dagger}[\phi] \right)=0,\,\,\,\,x\lesssim y.
\end{equation}
The S-matrix can be written as a Dyson expansion \cite{BS}
with respect to the couplings $k(x)$ considered as spacetime fields, i.e.,
\begin{equation} 
\label{Sexp}
\mathcal{S}[k(x)]=I+\sum_{n=1}^{\infty}\frac{1}{n!}\int 
{\rm d}x_{1}...{\rm d}x_{n}T\{{\cal S}_{n}(x_{1},...,x_{n})k(x_{1})...k(x_{n})\}.
\end{equation}
The conditions (\ref{Smatrix}) and (\ref{transl}) can be also expressed in terms 
of a functional Legendre transform mapping $\phi(x)$ into $k(x)$, leading to exactly 
the same expressions but in the variables $k(x)$. 
 
By setting $k(x)$ equal to a constant $k$, we revert to the usual couplings
and we can insert the expansion (\ref{Sexp}) into (\ref{Smatrix}). 
We obtain the following recursive relations for perturbation theory \cite{Joglekar1}:
\begin{equation} 
\label{norder}
{\mathcal C}_{n}={\rm i}{\cal S}_{n+1}(y,x_{1},...,x_{n})+{\rm i}\sum_{0\leq k \leq n-1}
\mathcal{P}\{{\cal S}_{k+1}(y,x_{1},..,x_{k}){\cal S}_{n-k}^{\dagger}(x_{k+1},...,x_{n})\} ,
\end{equation}
and 
\begin{equation} 
\label{norder2}
{\cal S}_{n}(x_{1},...,x_{n})+ {\cal S}^{\dagger}_{n}(x_{1},...,x_{n})
+\sum_{1\leq k \leq n-1}
{\mathcal P}\{{\cal S}_{k}(x_{1},..,x_{k}){\mathcal S}_{n-k}^{\dagger}(x_{k+1},...,x_{n})\}=0,
\end{equation}
where 
\begin{equation} 
\label{Sint}
{\mathcal S}_{n}=\int {\mathcal S}_{n}(x_{1},...,x_{n}){\rm d}x_{1}...{\rm d}x_{n},
\end{equation}
and $\mathcal{P}\{ \}$ is the sum over all partitions of $\{x_{1},...,x_{n}\}$ into
$k$ and $n-k$ elements. 
For example, the simplest is $\{x_{1},..,x_{k}\}$, $\{x_{k+1},..,x_{n}\}$. 

Thus, for the first two orders, the causality condition is described by \cite{Joglekar1} 
\begin{equation} 
\label{cc}
{\mathcal C}_{1}(x,y)={\rm i}\Bigr[{\cal S}_{2}(x,y)+{\cal S}_{1}(x){\cal S}_{1}^{\dagger}(y)\Bigr]=0,
\end{equation}
\begin{equation} 
\label{cc2}
{\mathcal C}_{2}(x,y)={\rm i}\Bigr[{\cal S}_{3}(x,y,z)+{\cal S}_{1}(x){\cal S}_{2}^{\dagger}(y,z)
+{\cal S}_{2}(x,y){\cal S}_{1}^{\dagger}(z)+{\cal S}_{2}(x,z){\cal S}_{1}^{\dagger}(y)\Bigr]=0.
\end{equation}
On the other hand, unitarity is expressed by \cite{Joglekar1}
\begin{equation} 
\label{un1}
{\mathcal S}_{1}(x)+{\mathcal S}_{1}^{\dagger}(x)=0,
\end{equation}
\begin{equation} 
\label{un2}
{\mathcal S}_{2}(x,y)+{\mathcal S}_{2}^{\dagger}(x,y)
+{\mathcal S}_{1}(x){\mathcal S}_{1}^{\dagger}(y)+{\mathcal S}_{1}(y){\mathcal S}_{1}^{\dagger}(x)=0.
\end{equation}
Relations (\ref{cc})-(\ref{cc2}) and (\ref{un1})-(\ref{un2}) can be used as a 
test of causality and unitarity to all orders of perturbation theory. 
For instance, by considering the integral relations corresponding to 
(\ref{cc})-(\ref{cc2}) one can write
\begin{equation} 
\label{C1}
\mathcal{C}_{1}=\int {\rm d}^{4}x {\rm d}^{4}y
[\theta(x_{0}-y_{0})\mathcal{C}_{1}(x,y)+\theta(y_{0}-x_{0})\mathcal{C}_{1}(y,x)]=0,
\end{equation}
\begin{equation} 
\label{C2}
\mathcal{C}_{2}=\int {\rm d}^{4}x {\rm d}^{4}y {\rm d}^{4}z 
\mathcal{C}_{2}(x,y,z)\theta(x_{0}-y_{0})\theta(y_{0}-z_{0})+ 5 \; {\rm symmetric}\, 
{\rm terms} =0.
\end{equation}
As a consequence, we will use the following criterion in the next section 
as a signal of causality violation: 
a {\it momentum dependence in $\mathcal{S}_{1}$}, implying a momentum dependence
in $\mathcal{C}_{1}$, is {\it surely a signal of causality violation and unitarity violation}. 
In fact, if the amplitude is proportional to a function of the Mandelstam 
variables $c^{n}/\Lambda^2$ with $n>1$ $(c=s,t,u)$ for $c>\Lambda^{2}$
a breakdown of unitarity and causality will occur, leading to an inconsistent theory
up to the scale $\Lambda$. In order to arrive to this conclusion, it is not 
important to perform Lehmann-Szymanszik-Zimmermann transformations
in these recursive conditions: the violations will be manifest by the momentum dependence. 

\section{Supergraphs and cancellations of divergences}

In this section, we study perturbation theory in 
supergraphs formalism rather than ordinary Feynman diagrams.
For an useful reference in the ordinary supersymmetric case, 
see Refs. \cite{NR1,susy1,susy2,susy3}
We would like to give prescriptions in order to calculate 
superfield Green's functions in a nonlocal model:
\begin{equation}
\label{Green}
<0|\mathcal{T}\left\{\Phi^{(1,2)}(z^{1})....\Phi^{(1,2)}(z^{r})
\Phi^{(1,2)^{\dagger}}(z^{r+1})...\Phi^{(1,2)^\dagger}(z^{s})\right\}|0>,
\end{equation}
where $z^{r}=(x^{r},\theta^{r},\bar{\theta}^{r})$ are superspace coordinates,
$\Phi^{(1)}$ and $\Phi^{(2)}$ are ordinary and auxiliary superfields, respectively.
The propagator is the basic building block of a generic perturbation theory,
and we can construct it by propagators of its component fields:
\begin{eqnarray}
\label{propagator1}
<0|\mathcal{T}\{\Phi^{(1,2)}(y,\theta)\Phi^{(1,2)}(y'\theta')\}|0>
&=& <0|\mathcal{T}\{[\phi^{(1,2)}(y)+\sqrt{2}\psi^{(1,2)}(y)\theta+F^{(1,2)}\theta\theta]
\nonumber \\
& \times & [\phi^{(1,2)}(y')+\sqrt{2}\psi^{(1,2)}(y')\theta'+F^{(1,2)}\theta'\theta'] \}|0>,
\end{eqnarray}
that we can expand, and the only nonvanishing terms are
\begin{eqnarray}
\label{dev1}
\; & \; &
\theta \theta<0|\mathcal{T}\{F^{(1,2)}(y)\phi^{(1,2)}(y')\}|0>
+\theta' \theta'<0|\mathcal{T}\{\phi^{(1,2)}(y)F^{(1,2)}(y')\}|0>
\nonumber \\
&+& 2\theta'^{\beta} \theta^{\alpha}<0|\mathcal{T}\{\psi^{(1,2)}_{\alpha}(y)\psi^{(1,2)}_{\beta}(y')\}|0>,
\end{eqnarray}
with $y,y^{\dagger}$ defined, as usual,
as $(y,y^{\dagger})=x\pm {\rm i} \theta \sigma \bar{\theta}$. The 
$\phi^{(1,2)},F^{(1,2)},\psi^{(1,2)}$ propagators contain $\Delta^{(1,2)}_{F}(x-x')$ 
in the form
\begin{equation}
\label{psi}
<0|\mathcal{T}\left\{\psi^{(1,2)}_{\alpha}(x)\psi^{(1,2)^{\beta}}(x') \right\}|0>
\equiv {\rm i} \delta_{\alpha}^{\beta}m\Delta^{(1,2)}_{F}(x-x'),
\end{equation}
\begin{equation}
\label{psi2}
<0|\mathcal{T}\left\{\bar{\psi}^{(1,2)^{\dot{\alpha}}}(x)\bar{\psi}^{(1,2)}_{\dot{\beta}}(x') 
\right\}|0> \equiv {\rm i}\delta^{\dot{\alpha}}_{\dot{\beta}}m\Delta^{(1,2)}_{F}(x-x'),
\end{equation}
\begin{equation}
\label{psi3}
<0|\mathcal{T}\left\{\psi^{(1,2)}_{\alpha}(x)\bar{\psi}^{(1,2)}_{\dot{\beta}}(x')\right\}|0>
\equiv \sigma_{\alpha\dot{\beta}}^{\mu}\partial_{\mu}\Delta_{F}^{(1,2)}(x-x'),
\end{equation}
\begin{equation}
\label{phi}
<0|\mathcal{T}\left\{\phi^{(1,2)}(x)\phi^{(1,2)^{*}}(x') \right\}|0>
\equiv {\rm i}\Delta_{F}^{(1,2)}(x-x'),
\end{equation}
\begin{equation}
\label{AF}
<0|\mathcal{T}\left\{\phi^{(1,2)}(x)F^{(1,2)}(x')\right\}|0> 
\equiv -{\rm i}m\Delta_{F}^{(1,2)}(x-x'),
\end{equation}
\begin{equation}
\label{AF1}
<0|\mathcal{T}\left\{\phi^{(1,2)^{*}}(x)F^{(1,2)^{*}}(x')\right\}|0>
\equiv -{\rm i}m\Delta_{F}^{(1,2)}(x-x'),
\end{equation}
\begin{equation}
\label{FF}
<0|\mathcal{T}\left\{ F^{(1,2)}(x)F^{(1,2)^{*}}(x')\right\}|0>
\equiv {\rm i}\Box \Delta_{F}^{(1,2)}(x-x').
\end{equation}
By substituting these definitions of propagators into (\ref{dev1}), we obtain 
$-{\rm i}m\Delta_{F}^{(1,2)}(y-y')(\theta-\theta')^{2}$, whereas 
\begin{equation}
\label{pr1}
<0|\mathcal{T}\left\{\Phi^{(1,2)}(x,\theta,\bar{\theta})\Phi^{(1,2)}(x',\theta',\bar{\theta}') \right\}|0>
\equiv -{\rm i}m \delta(\theta-\theta')\rm exp\left[i(\theta \sigma^{\mu}\bar{\theta}
-\theta' \sigma^{\mu}\bar{\theta}')\partial_{\mu}\right]\Delta_{F}^{(1,2)}(x-x'),
\end{equation}
\begin{eqnarray}
\label{pr2}
\; & \; & 
<0|\mathcal{T}\left\{\Phi^{(1,2)^{\dagger}}(x,\theta,\bar{\theta})
\Phi^{(1,2)^{\dagger}}(x',\theta',\bar{\theta}') \right\}|0> 
\nonumber \\
& \equiv & -{\rm i} m \delta(\theta-\theta')\rm exp\left[-i(\theta \sigma^{\mu}\bar{\theta}
-\theta' \sigma^{\mu}\bar{\theta}')\partial_{\mu}\right]\Delta^{(1,2)}_{F}(x-x'),
\end{eqnarray}
and similarly we can construct
\begin{equation}
\label{pr3}
<0|\mathcal{T}\left\{\Phi^{(1,2)}(x.\theta,\bar{\theta})\Phi^{(1,2)^{\dagger}}
(x'.\theta',\bar{\theta}') \right\}|0>
\equiv {\rm i}{\rm exp}\left[ {\rm i}(\theta \sigma^{\mu}\bar{\theta}+\theta'\sigma^{\mu}\bar{\theta}'
-2\theta \sigma^{\mu}\bar{\theta}')\partial_{\mu}\right]\Delta_{F}^{(1,2)}(x-x').
\end{equation}

The two-point functions $\Delta_{F}^{(1,2)}$ read as
\begin{equation} 
\label{function1}
\Delta_{F}^{(1)}=\frac{1}{(\Box-m^{2})}{\rm e}^{\frac{(\Box^{2}-m^{2})}{\Lambda^{2}}},
\end{equation}
\begin{equation} 
\label{function2}
\Delta_{F}^{(2)}=\frac{1}{(\Box-m^{2})}\left(I-{\rm e}^{\frac{(\Box^{2}-m^{2})}{\Lambda^{2}}}\right),
\end{equation}
according to definitions given in Section II A.
Note that (\ref{function1},\ref{function2}) are
encoding not only an usual free-field propagator,
but an infinite number of derivatives resulting from 
higher order powers of the Kahler terms $\int {\rm d}^{2}\theta {\rm d}^{2}\bar{\theta}
\mathcal{K}^{(1,2)^{ab}}\Phi_{_{a}}^{(1,2)^{\dagger}}\Phi^{(1,2)}_{_{b}}$, i.e., 
$\phi^{(1,2)*} \Box^{n}\phi^{(1,2)}$ and $\bar{\psi}^{(1,2)} (\sigma_{\mu}\partial^{\mu})^{n}\psi^{(1,2)}$
in terms of scalar and fermion components of the chiral field.
In a nonlocal model, it is more natural and simpler 
to consider these higher-derivative interactions 
just in effective smeared propagators,
as (\ref{function1},\ref{function2}). 

With these prescriptions for propagators, we can evaluate 
all superfields' Green functions, to any order of perturbation theory.
The $n$-th order contribution is 
\begin{equation} 
\label{intc1}
<0|\mathcal{T}\left\{ \Phi(z^{1})...\Phi^{\dagger}(z^{r+1})...\int \mathcal{L}_{{\rm int}}(x'_{1})
{\rm d}^{4}x'_{1}...\int \mathcal{L}_{{\rm int}}(x'_{n}){\rm d}^{4}x'_{n}\right\}|0>,
\end{equation}
with 
\begin{equation}
\label{intL}
\mathcal{L}_{{\rm int}}=\int {\rm d}^{2}\theta_{n}{\rm d}^{2}\bar{\theta}_{n}^{2}\frac{1}{3}
\left[g\Phi^{3}(x'_{n},\theta_{n},\bar{\theta}_{n})\delta(\bar{\theta}_{n})
+g^{*}(\Phi^{\dagger}(x'_{n},\theta_{n},\bar{\theta}_{n}))^{3}\delta(\bar{\theta}_{n})\right],
\end{equation}
that we can evaluate through Feynman diagrams, using Wick's theorem.
In the following subsections, we will show explicit 
applications of this perturbative machinery 
to a nonlocal supersymmetric model.

Superfields' Green functions are related to the generating functional $Z[J]$ by 
\begin{eqnarray}
\label{Gfunction}
\; & \; &
\mathcal{G}^{n}(z^{1},...,z^{m};z^{m+1},....,z^{n})
\nonumber \\
&=& (-{\rm i})^{n}\left[\frac{\delta}
{\delta J^{(1,2)}(z^{1})}...\frac{\delta}{\delta J^{(1,2)}(z^{m})}\frac{\delta}
{\delta J^{(1,2)^{\dagger}}(z^{m+1})}...\frac{\delta}{\delta J^{(1,2)^{\dagger}}(z^{n})}
Z[J^{(1,2)},J^{(1,2)^{\dagger}}]\right]_{J^{(1,2)}=J^{(1,2)^{\dagger}}=0},
\end{eqnarray}
where 
\begin{equation}
\label{ZJJdag}
Z[J^{(1,2)},J^{(1,2)^{\dagger}}]={\rm exp}\left\{{\rm i}\int d^{4}x\mathcal{L}_{{\rm int}}
\left(\frac{\delta}{\delta J^{(1,2)}},\frac{\delta}{\delta J^{(1,2)^{\dagger}}} 
\right) \right\}Z_{0}[J^{(1,2)},J^{(1,2)^{\dagger}}],
\end{equation}
\begin{equation}
\label{ZJzero}
Z_{0}[J^{(1,2)},J^{(1,2)^{\dagger}}]={\rm exp}\left[ -\frac{{\rm i}}{2}\int {\rm d}^{4}x {\rm d}^{2}\theta {\rm d}^{4}x' 
{\rm d}^{4}\theta' (J^{(1,2)}(z),J^{(1,2)^{\dagger}}(z))
\Delta^{(1,2}_{NL}(J^{(1,2)}(z),J^{(1,2)^{\dagger}}(z))^{T}\right],
\end{equation}
\begin{equation}
\label{PropagatorG}
\Delta_{NL}^{(1)}=\frac{1}{(\Box-m^{2})}{\rm e}^{\frac{\Box-m^{2}}{\Lambda^{2}}} \left( \begin{array}{cc} 
\frac{D^{2}}{\Box} & I
\ \\ I & \frac{\bar{D}^{2}}{\Box} \ \\
\end{array} \right) \delta(z-z'),
\end{equation}
\begin{equation}
\label{PropagatorG2}
\Delta_{NL}^{(2)}=\frac{1}{(\Box-m^{2})}\left[I-{\rm e}^{\frac{\Box-m^{2}}{\Lambda^{2}}}\right] \left( \begin{array}{cc} 
\frac{D^{2}}{\Box} & I
\ \\ I & \frac{\bar{D}^{2}}{\Box} \ \\
\end{array} \right) \delta(z-z').
\end{equation}

Now we are ready to display Feynman diagrams 
for supergraphs in nonlocal Wess-Zumino, according to the following recipe:
\vskip 0.3cm
\noindent
(i) Write a chiral field $\Phi_{1}$ (in-going) or $\bar{\Phi}$ (out-going) 
for any external line.
\vskip 0.3cm
\noindent
(ii) For each vertex $\Phi^{(1,2)}\Phi^{(1,2)}\Phi^{(1,2)}$
(and $\bar{\Phi}^{(1,2)}\bar{\Phi}^{(1,2)}\bar{\Phi}^{(1,2)}$)
 write a $-\frac{1}{4}\bar{D}^{2}$ ($\frac{1}{4}D^{2}$) acting 
on one internal propagator.
\vskip 0.3cm
\noindent
(iii) Write $\frac{1}{3}g$ (couplings) for each vertex
($\frac{1}{3}g^{*}$ for a vertex of antichiral fields' interactions).  
\vskip 0.3cm
\noindent
(iv) Use propagators (\ref{PropagatorG}) for each 
internal line of ordinary chiral (antichiral fields) 
and (\ref{PropagatorG2}) for each internal line
of auxiliary fields.
\vskip 0.3cm
\noindent
(v) Compute the combinatory factor and integrate for each vertex 
as $\int {\rm d}^{2}\theta {\rm d}^{4}x$ (in the antichiral case write instead ${\rm d}^{2}\bar{\theta}$).

In the following sections, we will apply to superfield propagators our technique for evaluation of 
radiative corrections, couplings and tadpoles.
Eventually, we will consider scattering processes.  

\subsection{Radiative corrections to superfield propagators} 

Let us consider one-loop radiative corrections 
to superfield two-point functions.
The possible diagrams in a nonlocal Wess-Zumino model are shown in Fig. 1.
\begin{figure}
\includegraphics[scale=0.08]{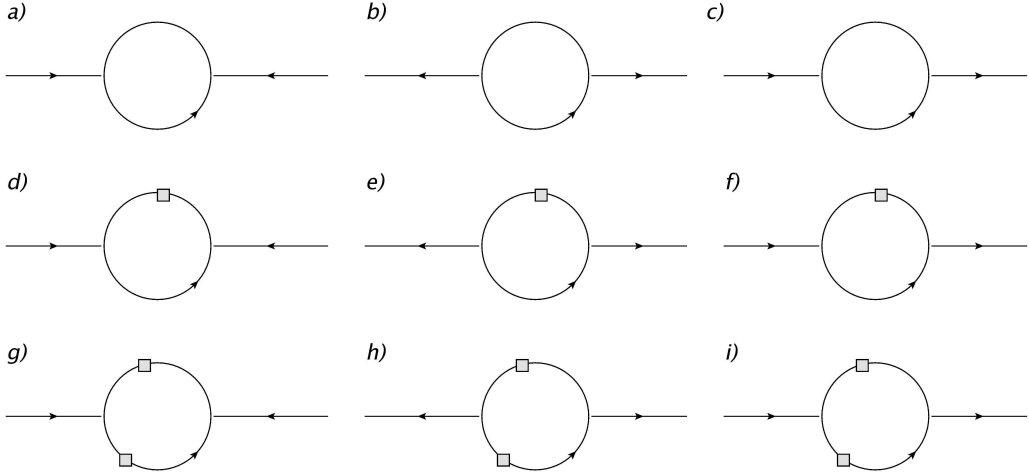} 
\caption{Radiative corrections to the two-point functions (superfield propagators)
of $\Phi^{(1)}\Phi^{(1)}$, $\Phi^{(1)^{\dagger}}\Phi^{(1)^{\dagger}}$ and $\Phi^{(1)}\Phi^{(1)^{\dagger}}$.
Diagrams (a)-(b)-(c) are radiative corrections to  $\Phi^{(1)}\Phi^{(1)}$
two-point functions, with contributions resulting not only from $\Phi^{(1)}\Phi^{(1)}$ 
internal lines, but also from auxiliary fields $\Phi^{(2)}\Phi^{(2)}$.
Diagrams (d)-(e)-(f) are corrections to  $\Phi^{(1)^{\dagger}}\Phi^{(1)^{\dagger}}$ with 
internal lines $\Phi^{(1,2)^{\dagger}}\Phi^{(1,2)^{\dagger}}$. 
Diagrams (g)-(h)-(i) are corrections to $\Phi^{(1)}\Phi^{(1)^{\dagger}}$
by internal lines $\Phi^{(1,2)}\Phi^{(1,2)^{\dagger}}$.
The only diagrams different from zero are (g)-(h)-(i),
the other ones are all automatically vanishing.  }
\label{plot}  
\end{figure}
All integrals (a)-(b)-(c)-(d)-(e)-(f) are automatically vanishing:
vertices in (a)-(b)-(c) are proportional to 
$\delta^{2}(\theta-\theta')=\delta(0)$ which is set to $0$ in dimensional regularization;
vertices in (d)-(e)-(f) are proportional to 
${\rm d}^{2}\bar{\theta}$ and ${\rm d}^{2}\bar{\theta}'$, {\it i.e.}, they are proportional to 
$\delta^{2}(\bar{\theta}-\bar{\theta}')=\delta(0)$ which is set to $0$ with the same understanding.
As a consequence, we obtain the following 
important result, that can be simply generalized 
to $n$ loops: {\it all contributions to mass renormalization
are vanishing also in nonlocal supersymmetric models}.
Let us now consider the contributions of (g)-(h)-(i) in Fig. 1:
\begin{eqnarray}
\label{Ig}
\mathcal{I}_{g}&=& \int {\rm d}^{4}x {\rm d}^{4}x'{\rm d}^{2}\theta {\rm d}^{2}\theta' {\rm d}^{2}\bar{\theta}
{\rm d}^{2}\bar{\theta}'\delta(\bar{\theta})\delta(\theta')\Phi^{(1)}(x,\theta,\bar{\theta})
{\rm exp}[{\rm i}(\theta \sigma^{\mu}\bar{\theta}+\theta' \sigma^{\mu}\bar{\theta}'
-2\theta \sigma^{\mu}\bar{\theta}')\partial_{\mu}]
\nonumber \\
& \times & \Delta_{F}^{(1)}(x-x'){\rm exp}[{\rm i}(\theta \sigma^{\mu}\bar{\theta}+\theta'\sigma^{\mu}
\bar{\theta}'-2\theta\sigma^{\mu}\bar{\theta}')\partial_{\mu}]\Delta_{F}^{(1)}(x-x')
\Phi^{(1)^{\dagger}}(x',\theta',\bar{\theta}'),
\end{eqnarray}
\begin{eqnarray}
\label{Ih}
\mathcal{I}_{h}&=& \int {\rm d}^{4}x {\rm d}^{4}x' {\rm d}^{2}\theta {\rm d}^{2}\theta' 
{\rm d}^{2}\bar{\theta}{\rm d}^{2} \bar{\theta}'
\delta(\bar{\theta})\delta(\theta')\Phi^{(1)}(x,\theta,\bar{\theta})
{\rm exp}[{\rm i}(\theta \sigma^{\mu}\bar{\theta}+\theta' \sigma^{\mu}\bar{\theta}'
-2\theta \sigma^{\mu}\bar{\theta}')\partial_{\mu}],
\nonumber \\
& \times & \Delta_{F}^{(1)}(x-x'){\rm exp}[{\rm i}(\theta \sigma^{\mu}\bar{\theta}+\theta'\sigma^{\mu}
\bar{\theta}'-2\theta\sigma^{\mu}\bar{\theta}')\partial_{\mu}]\Delta_{F}^{(2)}(x-x')
\Phi^{(1)^{\dagger}}(x',\theta',\bar{\theta}'),
\end{eqnarray}
\begin{eqnarray}
\label{Ii}
\mathcal{I}_{i}&=& \int {\rm d}^{4}x {\rm d}^{4}x' {\rm d}^{2}\theta {\rm d}^{2}\theta' 
{\rm d}^{2}\bar{\theta}{\rm d}^{2}\bar{\theta}'
\delta(\bar{\theta})\delta(\theta')\Phi^{(1)}(x,\theta,\bar{\theta})
{\rm exp}[{\rm i}(\theta \sigma^{\mu}\bar{\theta}+\theta' \sigma^{\mu}\bar{\theta}'
-2\theta \sigma^{\mu}\bar{\theta}')\partial_{\mu}]
\nonumber \\
& \times & \Delta_{F}^{(2)}(x-x'){\rm exp}[{\rm i}(\theta \sigma^{\mu}\bar{\theta}+\theta'\sigma^{\mu}
\bar{\theta}'-2\theta\sigma^{\mu}\bar{\theta}')\partial_{\mu}]\Delta_{F}^{(2)}(x-x')
\Phi^{(1)^{\dagger}}(x',\theta',\bar{\theta}').
\end{eqnarray}

These integrals are contributing to the corrections of the wave functions. 
In the limit of $\Lambda \rightarrow \infty$, these integrals give logaritmic
divergences, that can be absorbed in the wave function renormalization. 
In this case, we are integrating extra exponential factors 
of the momenta, suppressing divergences in (\ref{Ig}), 
but reintroducing {\it wild divergences} ({\it i.e.}, out of control) in the integral (\ref{Ii}). 
In particular, an infinite power series of momenta is expected,
leading to a nonrenormalizable theory. 
However, these extra corrections $p^{2n}$ are suppressed as $\Lambda^{-2n}$ at the 
$n$-th order. Thus, nonrenormalizzable contributions 
can be handled and controlled if $\Lambda>>\bar{E}$,
where $\bar{E}$ is the energy scale observed in laboratories. 
Of course, nonrenormalizzable contributions are expected 
to be controlled by a theory beyond a nonlocal quantum 
field model, bearing in mind string theory as a natural completion of our model.  
Anyway, the importance of supersymmetry is manifest 
in the automatic cancellations of infinitely many infinities 
originating from Figs. 1-(a)-(b)-(c)-(d)-(e)-(f).

\subsection{Tadpoles and corrections to coupling constants}

In this section, we consider possible tadpoles contributions 
and corrections to coupling constants, in our nonlocal model. 
Without supersymmetry, it is generically expected that 
these contributions are wildly divergent 
as an infinite power series of the momenta. We show all relevant contributions in Fig. 2.
However, in supersymmetry, supergraphs' formalism 
shows manifestly that all tadpoles cancel each other, because
all diagrams are proportional to $\delta(0)$.
As a consequence, we obtain again a cancellation 
of an infinite number of infinities 
just by virtue of supersymmetry. In this case, there are no other possible 
tadpoles not cancelled by supersymmetry.

\begin{figure}
\includegraphics[scale=0.13]{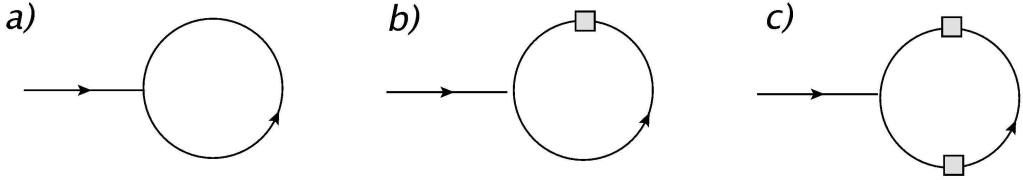} 
\caption{Tadpole diagrams resulting from $\Phi^{(1,2)}$ (ordinary and auxiliary fields).}
\label{plot}  
\end{figure}

Similarly, also possible corrections to 
superpotential terms $\Phi^{3}$ (Fig.3) and $\bar{\Phi}^{3}$
are identically zero because they are again proportional to $\delta(0)$.

\begin{figure}
\includegraphics[scale=0.08]{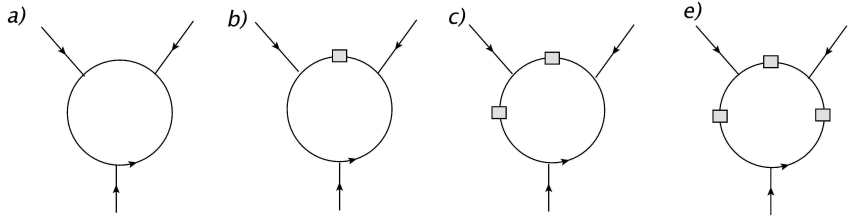} 
\caption{Corrections to $\Phi^{3}$ 
originating from $\Phi^{(1,2)}$ (ordinary and auxiliary fields).
All these diagrams are vanishing. Similarly, also 
diagrams with all arrows going out $(\Phi^{\dagger})^{3}$ are automatically equal to zero.}
\label{plot}  
\end{figure}

\section{One-loop acausal diagrams in $\phi\phi \rightarrow \phi\phi$ scattering}

In this section, we reconsider the same acausal one-loop process in a $\lambda \left(\phi^{(1)}\right)^{4}$ scalar 
self-interacting model, evaluated in Ref. \cite{Joglekar1}. As shown in this reference, 
the scattering $\phi^{(1)}\phi^{(1)} \rightarrow \phi^{(1)}\phi^{(1)}$
at one loop has a divergent amplitude (see Fig. 4). 
The harmful contribution results from the auxiliary field $\phi^{(2)}$ propagating inside the 
loops (off-shell) and interacting with $\phi^{(1)}$ as interaction terms $\sim \left(\phi^{(1)}\right)^{3}\phi^{(2)}$. 
By assuming the massless case $m=0$, the (renormalized) amplitude is \cite{Joglekar1,Joglekar2}
\begin{equation} 
\label{Amplitude1}
\mathcal{A}(s,t,u)_{{\rm Boson}}=\frac{9\lambda^{2}}{4\pi^{2}}\sum_{c=s,t,u}\sum_{n=0}^{\infty}
\frac{\left(\frac{c}{\Lambda^{2}} \right)^{n}(1-\frac{1}{2^{n}})}{n((n+1)!)},
\end{equation}
where $c=s,t,u$ are the Mandelstam variables, and we are summing over the three channels. 
We express this as a power-series expansion, because of the nonlinear functions on the vertices. 

As usual in quantum field theory, we can choose the Schwinger parametrization 
and we can rewrite the amplitudes as
the following integrals for the $s$-channel bosonic part (see Ref. \cite{Joglekar1,Joglekar2}): 
\begin{equation} 
\label{boson}
\mathcal{A}(s)_{{\rm B}}=\frac{9\lambda^{2}}{4\pi^{2}}\int_{0}^{1/2}{\rm d}x
\int_{\frac{1}{(1-x)}}^{\frac{1}{x}}\frac{{\rm d}\zeta}{\zeta}{\rm exp}
\left\{- \frac{\zeta}{\Lambda^{2}}(m^{2}-x(1-x)s) \right\} .
\end{equation}
The asymptotic expansion of this integral in the neighboorhood of $s=0$ reads as 
\begin{equation}
\label{Amplitudeesp}
\mathcal{A}_{B}(s) \sim \sum_{n=0}^{2}a_{n}(m,\Lambda)s^{n}+{\rm O}(s^{3}),
\end{equation}
where the coefficients $a_{0,1,2}$ are given by 
\begin{equation} 
\label{0th}
a_{0}(m,\Lambda)=\frac{9\lambda^{2}}{4\pi^{2}}\int_{0}^{1/2}{\rm d}x\int_{\frac{1}{(1-x)}}^{1 \over x}
\frac{{\rm d}\zeta}{\zeta}{\rm e}^{-\frac{m^{2}\zeta}{\Lambda^{2}}},
\end{equation}
\begin{equation} 
\label{1th}
a_{1}(m,\Lambda)=\frac{9\lambda^{2}}{4\pi^{2}}\int_{0}^{1/2}{\rm d}x\int_{\frac{1}{(1-x)}}^{1 \over x}
{\rm d}\zeta \frac{x(1-x)}{\Lambda^{2}}{\rm e}^{-\frac{m^{2}\zeta}{\Lambda^{2}}},
\end{equation}
\begin{equation} 
\label{2th}
a_{2}(m,\Lambda)=\frac{9\lambda^{2}}{4\pi^{2}}\int_{0}^{1/2}{\rm d}x\int_{\frac{1}{(1-x)}}^{1 \over x}
\zeta {\rm d}\zeta \frac{x^{2}(1-x)^{2}}{2\Lambda^{4}}{\rm e}^{-\frac{m^{2}\zeta}{\Lambda^{2}}}.
\end{equation}
As we said for the massless case, the zeroth and first order 
of the expansion (\ref{Amplitudeesp})
are constants cancelled in the renormalization (subtraction of counterterms). 
The result can be rewritten as
\begin{equation} 
\label{rewa2}
\mathcal{A}_{B}(s,t,u)=\frac{9\lambda^{2}}{4\pi^{2}}
\sum_{c=s,t,u}\left[ -{\rm log}[c/\Lambda^{2}]+
\sum_{n=0}^{\infty}\frac{1-2^{-n}}{(n+1)!n}\left(\frac{c}{\Lambda^{2}}\right)^{n}\right],
\end{equation}
where the first logarithmic contributions are the analogous of 
the local QFT result. 

\begin{figure}
\includegraphics[scale=0.80]{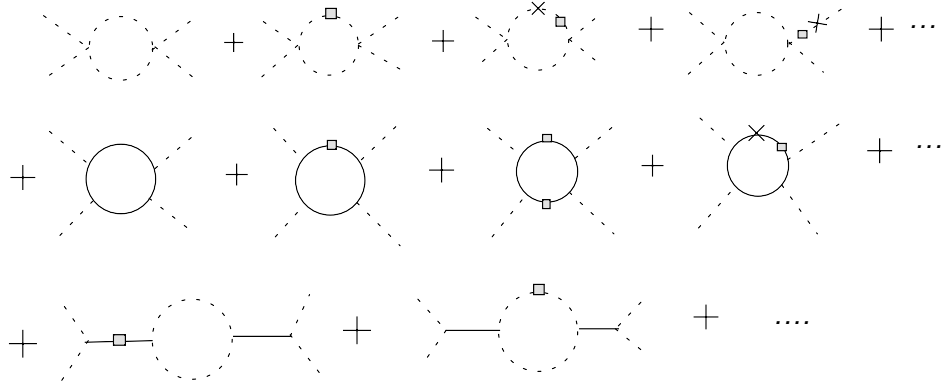} 
\caption{$\phi^{1}\phi^{1} \rightarrow \phi^{1}\phi^{1}$: One-loop diagrams leading to violations of causality. 
Dashed lines represent scalars $\phi^{1,2}$, continuous lines describe 
fermions $\psi^{1,2}$. Boxes on the lines represent auxiliary fields $\phi^{2}$ and $\psi^{2}$ 
(dashed and continuous, respectively). Bilinear mixing terms between 
standard fields and auxiliary ones are represented as crosses. Other, trivial complications
of these diagrams are not reported. 
By virtue of supersymmetry, all diagrams, with all entering arrows, cancel each other.}
\label{plot}  
\end{figure}

We also report evaluation of the diagram in Fig. 4 (the first one therein),
calculated in \cite{Joglekar3}:
\begin{equation}
\label{amplitude2}
\mathcal{A}_{B}(s,t,u)=\frac{9\lambda^{2}}{8\pi^{2}}\sum_{c=s,t,u}
\left[-{\rm log}\frac{c}{\Lambda^{2}}-2\sum_{n=1}^{\infty}\frac{1}{n! (n+1)}
\left(1-\frac{1}{2^{n+1}} \right)\left(\frac{c}{\Lambda^{2}}\right)^{n}\right],
\end{equation}
where the first logarithmic contributions are the analogous of 
the local quantum field theory result. 

However, in a nonlocal Wess-Zumino model, we have to consider 
also other diagrams resulting from the fermionic sector. 
In fact, we have to consider carefully all possible one-loop diagrams 
involving also all interactions between auxiliary superfields and standard superfields:
Yukawa like terms $\phi^{(2)}\bar{\psi}_{L}^{(1)}\psi_{R}^{(1)}+{\rm h.c.}$, 
$\phi^{(1)}\bar{\psi}_{L}^{(2)}\psi_{R}^{(1)}+{\rm h.c.}$; interactions in the scalars' sector 
$\left(\phi^{(1)}\right)^{3} \left(\phi^{(2)}\right)$ and $\left(\phi^{(1)}\right)^{2} \left(\phi^{(2)}\right)^{2}$ 
etc., mass-mixing terms $m\bar{\psi}^{(1)}_{L}\psi^{(2)}+{\rm h.c.}$ etc.
The computation might become quite tedious if we were to consider 
all numerical prefactors in the Taylor power series expansions 
inside the integrals and the complications from trace operations, 
the presence of particles' masses etc. 
But we would like to note that all parameters in the diagrams are connected by supersymmetry,
and the fermionic contributions are opposite in sign with respect to the bosonic ones. 
Thus, from a naive power-counting, it seems reasonable that some cancellations
will occur in this case. 
Now let us switch our language from ordinary Feynman diagrams to supergraphs. 
In Fig.5, relevant one-loop supergraphs are reported. 
The evaluation of all these diagrams seems unfeasible.
However, we can conclude that 
all diagrams in Fig. 5, except (d), are equal to zero
because all these diagrams are proportional 
to $\delta(0)$, resulting from internal 
$\Phi^{1,2}\Phi^{1,2}$ and $\bar{\Phi}^{1,2}\bar{\Phi}^{1,2}$.
In fact, we have the following amplitudes:
\begin{eqnarray}
\label{firsephiphi1}
\mathcal{A}_{1}&=& C_{1}\frac{g^{4}}{12^{4}}\int \int \int \int {\rm d}^{4}x_{1,2,3,4}
{\rm d}^{2}\theta_{1,2,3,4}{\rm d}\Phi(z_{1})\Phi(z_{2})(\bar{D}^{2})\left[\frac{1}{(\Box-m^{2})}
{\rm e}^{\frac{\Box-m^{2}}{\Lambda^{2}}}\right]_{z_{1},z_{2}}
\nonumber \\
& \times & (\bar{D}^{2})\left[\frac{1}{(\Box-m^{2})}{\rm e}^{\frac{\Box-m^{2}}{\Lambda^{2}}}\right]_{z_{2},z_{3}}
(\bar{D}^{2})\left[\frac{1}{(\Box-m^{2})}{\rm e}^{\frac{\Box-m^{2}}{\Lambda^{2}}}\right]_{z_{3},z_{4}}
\nonumber \\
& \times & (\bar{D}^{2})\left[\frac{1}{(\Box-m^{2})}{\rm e}^{\frac{\Box-m^{2}}
{\Lambda^{2}}}\right]_{z_{4},z_{1}}\Phi(z_{3})\Phi(z_{4}) 
\nonumber \\
& \times & \delta(z_{1}-z_{2})\delta(z_{2}-z_{3})\delta(z_{3}-z_{4})\delta(z_{4}-z_{1}),
\end{eqnarray}
\begin{eqnarray}
\label{firsephiphi2}
\mathcal{A}_{2}&=& C_{2}\frac{g^{4}}{12^{4}}\int\int\int\int  {\rm d}^{4}x_{1,2,3,4}{\rm d}^{2}\theta_{1,2,3,4}
\Phi(z_{1})\Phi(z_{2})(\bar{D}^{2})\left[\frac{1}{(\Box-m^{2})}
\left(I-{\rm e}^{\frac{\Box-m^{2}}{\Lambda^{2}}}\right)\right]_{z_{1},z_{2}}
\nonumber \\
& \times & (\bar{D}^{2})\left[\frac{1}{(\Box-m^{2})}{\rm e}^{\frac{\Box-m^{2}}{\Lambda^{2}}}\right]_{z_{2},z_{3}}
(\bar{D}^{2})\left[\frac{1}{(\Box-m^{2})}{\rm e}^{\frac{\Box-m^{2}}{\Lambda^{2}}}\right]_{z_{3},z_{4}}
\nonumber \\
& \times & (\bar{D}^{2})\left[\frac{1}{(\Box-m^{2})}{\rm e}^{\frac{\Box-m^{2}}{\Lambda^{2}}}\right]_{z_{4},z_{1}}
\Phi(z_{3})\Phi(z_{4})
\nonumber \\
& \times & \delta(z_{1}-z_{2})\delta(z_{2}-z_{3})\delta(z_{3}-z_{4})\delta(z_{4}-z_{1}),
\end{eqnarray}
\begin{eqnarray}
\label{firsephiphi3}
\mathcal{A}_{3}&=& C_{3}\frac{g^{4}}{12^{4}}\int \int \int \int {\rm d}^{4}x_{1,2,3,4}{\rm d}^{2}\theta_{1,2,3,4}
\Phi(z_{1})\Phi(z_{2})(\bar{D}^{2})\left[\frac{1}{(\Box-m^{2})}
\left(I-{\rm e}^{\frac{\Box-m^{2}}{\Lambda^{2}}}\right)\right]_{z_{1},z_{2}}
\nonumber \\
& \times & (\bar{D}^{2})\left[\frac{1}{(\Box-m^{2})}
\left(I-{\rm e}^{\frac{\Box-m^{2}}{\Lambda^{2}}}\right)\right]_{z_{2},z_{3}}
(\bar{D}^{2})\left[\frac{1}{(\Box-m^{2})}{\rm e}^{\frac{\Box-m^{2}}{\Lambda^{2}}}\right]_{z_{3},z_{4}}
\nonumber \\
& \times & (\bar{D}^{2})\left[\frac{1}{(\Box-m^{2})}
{\rm e}^{\frac{\Box-m^{2}}{\Lambda^{2}}}\right]_{z_{4},z_{1}}\Phi(z_{3})\Phi(z_{4})
\nonumber \\
& \times & \delta(z_{1}-z_{2})\delta(z_{2}-z_{3})\delta(z_{3}-z_{4})\delta(z_{4}-z_{1}),
\end{eqnarray}
\begin{eqnarray}
\label{firsephiphi4}
\mathcal{A}_{4}&=& C_{4}\frac{g^{4}}{12^{4}}\int\int\int \int {\rm d}^{4}x_{1,2,3,4}{\rm d}^{2}\theta_{1,2,3,4}
\Phi(z_{1})\Phi(z_{2})(\bar{D}^{2})\left[\frac{1}{(\Box-m^{2})}
\left(I-{\rm e}^{\frac{\Box-m^{2}}{\Lambda^{2}}}\right)\right]_{z_{1},z_{2}}
\nonumber \\
& \times & (\bar{D}^{2})\left[\frac{1}{(\Box-m^{2})}
\left(I-{\rm e}^{\frac{\Box-m^{2}}{\Lambda^{2}}}\right)\right]_{z_{2},z_{3}}(\bar{D}^{2})
\left[\frac{1}{(\Box-m^{2})}\left(I-{\rm e}^{\frac{\Box-m^{2}}{\Lambda^{2}}}\right)\right]_{z_{3},z_{4}}
\nonumber \\
&\times &(\bar{D}^{2})\left[\frac{1}{(\Box-m^{2})}{\rm e}^{\frac{\Box-m^{2}}{\Lambda^{2}}}\right]_{z_{4},z_{1}}
\Phi(z_{3})\Phi(z_{4})
\nonumber \\
& \times & \delta(z_{1}-z_{2})\delta(z_{2}-z_{3})\delta(z_{3}-z_{4})\delta(z_{4}-z_{1}),
\end{eqnarray}
\begin{eqnarray}
\label{firsephiphi5}
\mathcal{A}_{5}&=& C_{5}\frac{g^{4}}{12^{4}}\int\int\int \int 
{\rm d}^{4}x_{1,2,3,4}{\rm d}^{2}\theta_{1,2,3,4}\Phi(z_{1})\Phi(z_{2})(\bar{D}^{2})
\left[\frac{1}{(\Box-m^{2})}\left(I-{\rm e}^{\frac{\Box-m^{2}}{\Lambda^{2}}}\right)\right]_{z_{1},z_{2}}
\nonumber \\
&\times & (\bar{D}^{2})\left[\frac{1}{(\Box-m^{2})}
\left(I-{\rm e}^{\frac{\Box-m^{2}}{\Lambda^{2}}}\right)\right]_{z_{2},z_{3}}(\bar{D}^{2})\left[\frac{1}{(\Box-m^{2})}
\left(I-{\rm e}^{\frac{\Box-m^{2}}{\Lambda^{2}}}\right)\right]_{z_{3},z_{4}}
\nonumber \\
& \times & (\bar{D}^{2})\left[\frac{1}{\Box-m^{2}}
\left(I-{\rm e}^{\frac{\Box-m^{2}}{\Lambda^{2}}}\right)\right]_{z_{4},z_{1}}
\Phi(z_{3})\Phi(z_{4})
\nonumber \\
& \times & \delta(z_{1}-z_{2})\delta(z_{2}-z_{3})\delta(z_{3}-z_{4})\delta(z_{4}-z_{1}).
\end{eqnarray}
Let us bear in mind the following useful relations:
\begin{equation}
\label{remindD1}
\frac{1}{16}\frac{\bar{D}^{2}D^{2}}{\Box}\Phi=\Phi,
\,\,\,\,\,\,\,\,\,\bar{D}\frac{1}{16}\frac{\bar{D}^{2}D^{2}}{\Box}\Phi=0,
\,\,\,\,\,\,\,\,{\rm d}^{2}\theta=-\frac{1}{4}\bar{D}^{2},
\end{equation}
the last being valid inside the integral $\int {\rm d}^{4}x$. 
$C_{1,2,3,4,5}$ are combinatorial factors. 
Amplitudes $\mathcal{A}_{6,7,8,9,10}$ in Fig. 5-(b)
have the same form of $\mathcal{A}_{1,2,3,4,5}$ respectively,
with suitable substitutions
$\Phi^{(1,2)}\rightarrow \Phi^{(1,2)^{\dagger}}$ and $\bar{D}\rightarrow D$.
In this form, cancellations are made explicit by the fact that 
the number of $\bar{D}$ (${\rm d}^{2}\theta$) is not 
equal to the number of $D$ (${\rm d}^{2}\bar{\theta}$) in these amplitudes, 
and this leads to $\delta(0)$s
inside the integral, as mentioned above. 
As far as $\mathcal{A}_{11,15}$ in Fig. 5-(c) is concerned,
as well as $\mathcal{A}_{11,15}$ but with reversed arrows (not reported in Fig. 5),
the number of $D$ is different by the one of $\bar{D}$,
so we have cancellations of these contributions. 
Note that among these contributions discussed here, 
there are not only $F$-terms, but also some possible $D$-terms
(three entering arrows, one going out; three going out, one entering;
two entering, two going out).
As a cascade, a $n$-loop generalization of this result is understood. 
We would like to remark that also in this case an infinite number of divergences
are automatically cancelled. This reflects the fact that 
fermionic and bosonic components, as the ones reported in Fig. 4 
(the possible contributions are not all written down in this figure) are canceling each other. 

Examples of cancellations just shown are
not only accidental properties of one-loop calculations, but they are 
expected at all orders of perturbation theory.
In fact, in a nonlocal supersymmetric model, 
the nonrenormalization theorem remains valid. 
This crucial point will be discussed in the next sections.
By virtue of these powerful properties 
we will conclude that all $F$-terms' processes discussed
above are not corrected by an infinite number of divergences. 
\begin{figure}
\includegraphics[scale=0.10]{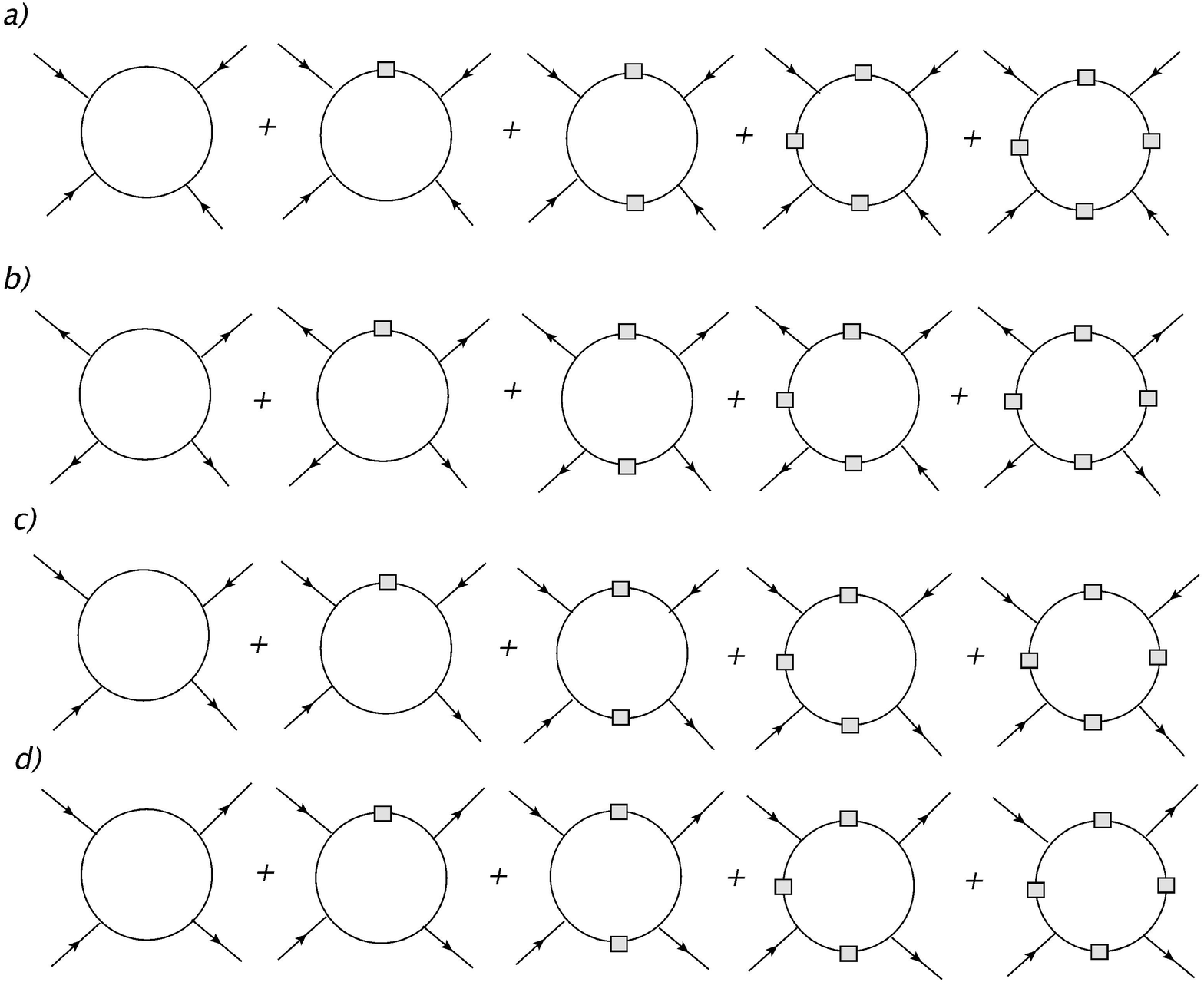} 
\caption{One-loop supergraphs contributions to $\phi^{(1)}\phi^{(1)}\rightarrow \phi^{(1)}\phi^{(1)}$. 
In these diagrams, propagators are all $\Phi^{(1,2)}\Phi^{(1,2)}$
in (a), all $\bar{\Phi}^{(1,2)}\bar{\Phi}^{(1,2)}$ in (b),
one $\Phi^{(1,2)}\Phi^{(1,2)}$ and three $\bar{\Phi}^{(1,2)}\Phi^{(1,2)}$ in (c); 
one $\Phi^{(1,2)}\Phi^{(1,2)}$, one $\bar{\Phi}^{(1,2)}\bar{\Phi}^{(1,2)}$ 
and two $\bar{\Phi}^{(1,2)}\Phi^{(1,2)}$ in (d).
By virtue of supersymmetry, all these diagrams are equal to zero, except (d).}
\label{plot}  
\end{figure}

\section{Acausality cancellations in $F$-terms to all orders of perturbation theory}

In the particular case studied in Sec. IV, 
we are considering the quantum one-loop corrections 
of a supersymmetric F-term. In fact, the interaction 
$\lambda (\phi^{1})^{4}$ is contained in the superpotential 
and obtained after the integration on the Grassmannian
(see Secs. IA and II). For this reason, we can say that the cancellations must 
occur to all orders of perturbation theory, by virtue of
the {\it nonrenormalization theorem} \cite{NR1,NR2} of supersymmetry (see Appendix). 
This tells us that in a supersymmetric theory
the cancellation of radiative corrections occurs to all orders,
and is a consequence of the holomorphic nature of the 
superpotential $\mathcal{W}(\Phi^{1}+\Phi^{2})$,
not permitting a perturbative renormalization 
of the supersymmetric $F$-terms. 
Thus, we have the following peculiar result: 
a nonlocal $\lambda \phi^{4}$ model is not consistent without supersymmetry, 
but with the help of supersymmetry we automatically 
obtain a consistent and perfectly computable 
example of a nonlocal supersymmetric quantum field theory without 
acausalities or other inconsistencies. 
Note that, in our nonlocal formulation, we have never modified the 
holomorphic superpotential, but only the K\"{a}hler potential (D-terms). 
For this reason, the nonrenormalization proof given by Seiberg \cite{NR2} is still valid.
In the next subsection, we summarize Seiberg's argument 
about the Wess-Zumino case with comments about 
its relations with locality\footnote{We would like to note that the nonlocal Wess-Zumino 
class of models under discussion is invariant under the superPoincar\'e group. 
As mentioned in the introduction, nonlocal theories violating the 
Poincar\'e invariance are not included in our considerations. For example, a Wess-Zumino model in 
noncommutative geometry is an example of a nonlocal supersymmetric 
quantum field theory, violating the local superPoincar\'e group. In this case, the nonrenormalization 
theorem does not protect the F-terms from quantum corrections. On the other hand, 
as shown in Ref. \cite{NCWZ}, there remains also in this case a residual supersymmetry ($\mathcal{N}=1/2$) 
protecting the F-terms from an infinite number of quantum corrections.}.

\subsection{Seiberg's argument about Wess-Zumino theory, relaxing locality}

Consider an initial (tree-level) superpotential  
(\ref{Wphii}). In the case $m=\lambda=0$,
we are restoring a global symmetry $G=U(1)\times U(1)\times U(1)_{R}$.
Such a global symmetry can be defined also in a nonlocal 
field theory in the same way of a local one. 
The field $\Phi$ transforms as $(1,1)$ under $G$. 
This constrains the dimensions of the couplings, i.e., ${\rm dim}(m) \in (-2,0)$ and 
${\rm dim}(\lambda) \in (-3,-1)$. 
A general superpotential invariant under $G$ can be written as 
\begin{equation} 
\label{WZexample}
\mathcal{W}_{{\rm eff}}=m\Phi^{2}f\left(\frac{\lambda \Phi}{m} \right),
\end{equation}
where $f$ is a generic  holomorphic function. 
We can always consider a power-series expansion of 
$f$. The $n$-th term invariant under the global group $G$ is 
$c_{n}\Phi^{n}$, where $c_{n}(\lambda,m)=\lambda^{n-2}\frac{1}{m^{n-3}}$. 
But this is also the same result obtained from a tree graph
with exchanges of the superfields $\Phi$. 
These have not to occur in the Wilsonian 
effective action; higher orders in $\lambda$ 
to the coefficient $c_{n}(\lambda,m)$ 
cannot be considered in the general structure of (\ref{WZexample}).
Thus, we conclude that the resulting effective superpotential is 
exactly equal to (\ref{WZexample}), and it is not renormalized. 

In general, we can ask ourselves whether, by relaxing the hypothesis of locality, 
this simple argument can be avoided. However, we want to stress
that in the nonlocal formulation of the Wess-Zumino model given in Sec. II, 
we have never modified the superpotentials, we have just deformed
the K\"{a}hler potentials. The argument is based only upon the holomorphic 
nature of the superpotentials, and in the nonlocal formulation this property is not affected. 
Thus, we conclude that Seiberg's argument remains valid also in 
a nonlocal Wess-Zumino model\footnote{A generalization to other cases given in Ref. \cite{NR2} can be considered.
For example, we can study a supersymmetric quantum chromodynamics
with a superpotential $S\bar{Q}Q+\lambda'S^{3}$,
with $S$ a gauge singlet of $U(1)_{S}$ (local gauge symmetry generalized 
as in section IIIA). Upon taking $\lambda=\lambda'=0$,
the theory is invariant under $G=SU(N_{f})_{L}\times SU(N_{f})_{R}\times U(1)_{V}\times U(1)_{S}\times U(1)_{R}$. 
We can then repeat an argument similar to the Wess-Zumino case, as shown in Ref. \cite{NR2}.}.

\subsection{Acausal corrections of $D$-terms in nonlocal Wess-Zumino}
\begin{figure}
\includegraphics[scale=0.10]{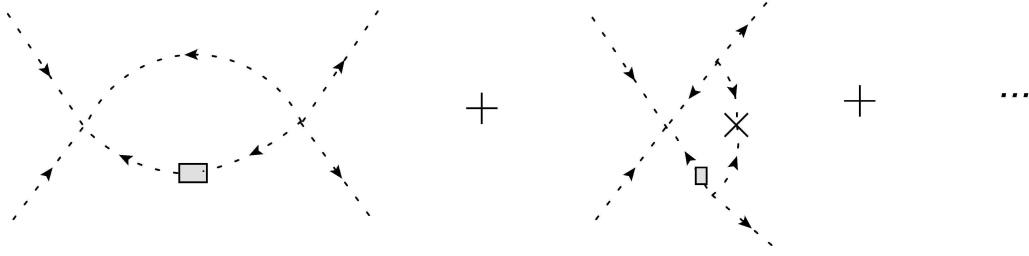} 
\caption{$\phi^{(1)}\phi^{(1)} \rightarrow \phi^{(1)}\phi^{(1)}$: we report two examples of one-loop 
diagrams not protected by the nonrenormalization theorem in nonlocal Wess-Zumino model. 
These are ordinary Feynman diagrams for the scalar sector.}
\label{plot}  
\end{figure}
In Wess-Zumino models, $F$-term superpotential terms are protected 
by acausal radiative corrections. We report contributions 
to $\phi\phi\rightarrow\phi\phi$.
Diagrams in Fig. 5-(d) lead to acausalities.
In Fig. 6, we give two examples of contributions 
from the scalar sector, with ordinary Feynman diagrams. 
These diagrams have divergences, violating causality 
and unitarity for $s,t,u>\Lambda$. 
The amplitude of the first one in Fig. 6, for example, is exactly equal 
to (\ref{boson}). As a consequence, not all possible acausal diagrams are cancelled, 
but infinitely many are cancelled just by virtue of supersymmetry, 
as discussed in the previous section. Consider the integral
\begin{eqnarray}
\label{firsephiphi}
\tilde{\mathcal{A}}_{2}&=& \tilde{C}_{2}\frac{g^{4}}{12^{4}}\int \int \int \int 
{\rm d}^{4}x_{1}{\rm d}^{2}\theta_{1}{\rm d}^{4}x_{2}{\rm d}^{2}\bar{\theta}_{2}
{\rm d}^{4}x_{3}{\rm d}^{2}\bar{\theta}_{3}  
{\rm d}^{4}x_{4}{\rm d}^{2}\theta_{4}
\Phi(z_{1})\bar{\Phi}(z_{2})\left[\frac{1}{(\Box-m^{2})}{\rm e}^{\frac{\Box-m^{2}}{\Lambda^{2}}}\right]_{z_{1},z_{2}}
\nonumber \\
& \times & D^{4}\bar{D}^{4}\left[\frac{1}{(\Box-m^{2})}{\rm e}^{\frac{\Box-m^{2}}
{\Lambda^{2}}}\right]_{z_{2},z_{3}}\left[\frac{1}{(\Box-m^{2})}
{\rm e}^{\frac{\Box-m^{2}}{\Lambda^{2}}}\right]_{z_{3},z_{4}} \left[\frac{1}
{(\Box-m^{2})}\left(I-{\rm e}^{\frac{\Box-m^{2}}{\Lambda^{2}}}\right)\right]_{z_{4},z_{1}}
\nonumber \\
&\times & \bar{\Phi}(z_{3})\Phi(z_{4})\delta(z_{1}-z_{2})
\delta(z_{2}-z_{3})\delta(z_{3}-z_{4})\delta(z_{4}-z_{1}).
\end{eqnarray}
In order to evaluate this integral, let us exploit the relations 
\begin{equation}
\label{rel1}
\int {\rm d}^{4}\theta_{n}\bar{D}^{2}D^{2}\delta(\theta_{n}-\theta_{1})
\delta(\bar{\theta}_{n}-\bar{\theta}_{1})=16\int {\rm d}^{4}\theta_{n}.
\end{equation}
Since we are not particularly interested in all numerical prefactors, 
in massless approximation, we note that
in momentum space this amplitude is approximated by
\begin{equation}
\label{divergence}
\tilde{\mathcal{A}}_{2}(p)\sim \int {\rm d}p \; p^{-5} 
{\rm e}^{-3p^{2}/\Lambda^{2}}(I-{\rm e}^{p^{2}/\Lambda^{2}}).
\end{equation}
From this we can argue that, even if there are 
divergent contributions, they will be suppressed
as $\Lambda^{-n}$ with $n>5$.
This result provides further evidence of the power of supersymmetry:
also acausal divergences become softer than in
nonsupersymmetric case by one order of magnitude. 
Thus, the amplitude calculated 
gets some cancellations by contributions  
of the fermionic sector. Not all divergences 
are cancelled, but up to fifth order in momenta, there are cancellations 
of $p^{n}$ with $n<5$ (logarithmic divergences remain).  
Other contributions, like ones with mass insertions, 
are more suppressed than (\ref{divergence}). 

\section{Nonlocal super Yang-Mills model}

We now want to consider a more realistic quantum field theory,
generalizing our observations for the Wess-Zumino case to a general supersymmetric Yang-Mills theory. 
For this purpose, we have to introduce a standard vector supermultiplet
$\{D^{1},A^{1}_{\mu}, \lambda^{1} \}$ and an auxiliary vector supermultiplet
$\{D^{2},A^{2}_{\mu}, \lambda^{2} \}$. The nonlocal action is 
\begin{eqnarray} 
\label{SYMnl}
\hat{S}[D^{1},A^{1},\lambda^{1}]&=& \frac{1}{2}\int {\rm d}^{4}x 
\left\{\hat{D}^{1}_{c} (a^{1})_{cd}\hat{D}_{d}^{1}
+\hat{A}^{1}_{c\mu} (b^{1})_{cd}^{\mu \nu} \hat{A}_{d\nu}^{1}-D^{2}_{c} (a^{2})_{cd}D_{d}^{2}
-\hat{A}^{2}_{c\mu}(b^{2})_{cd}^{\mu \nu} \hat{A}_{d\nu}^{2} \right.
\nonumber \\
&+& \left. \hat{\lambda}^{1}_{a\alpha}(f^{1})_{ab}^{\alpha \dot{\beta}}\hat{\lambda}^{1}_{b\dot{\beta}} 
- \lambda^{2}_{a\alpha}(f^{2})_{ab}^{\alpha \dot{\beta}}\lambda^{2}_{b\dot{\beta}}
+\mathcal{I}\{A^{1}+A^{2},\lambda^{1}+\lambda^{2}\}+{\rm h.c.}\right\} ,
\end{eqnarray}
where $D^{2},A^{2},\psi^{2}$ are functionals in the variables $D^{1},A^{1},\psi^{1}$ and solve the equations 
\begin{equation} 
\label{Funct}
\frac{\delta S}{\delta D^{2}}=\frac{\delta S}{\delta A^{2}}=\frac{\delta S}{\delta \lambda^{2}}=0,
\end{equation}
$a^{1},b^{1},f^{1}$ are the nonlocal operators containing masses and kinetic terms,
$a^{2},b^{2},f^{2}$ are the nonlocal operators of the auxiliary fields;
$a^{2},b^{2},f^{2}$ have exactly the same definitions given in Eq. (2.9)
and the redefined fields $\hat{D}^{1},\hat{A}^{1},\hat{\lambda}^{1}$ can be expressed as in Eqs. 
(\ref{phiat})-(\ref{psiat}). 

The nonlocal gauge symmetry of the action (\ref{SYMnl}) is as follows:
\begin{equation} 
\label{gsym}
\hat{\delta}_{\theta}(A^{1})_{a}^{\mu}=\mathcal{E}^{2}_{B}\{ -\theta_{a}^{,\mu}+g f_{bcd}(A^{1}_{c\mu}
+A^{2}_{c\mu}[A^{1},\lambda^{1}])\theta_{d}\},
\end{equation}
supplemented by the analogous transformation for the gaugino field $\lambda^{1}$.
The nonlocality in infinitesimal gauge transformations like (\ref{gsym})
results from the functional $A_{2}[D^{1},A_{1},\lambda_{1}]$. 
We can reformulate the action (\ref{SYMnl}) in a form explicitly invariant under supersymmetry. 
For simplicity, we do not consider Fayet-Iliopoulos terms,
that can be inserted in a well understood way. Thus, the supersymmetric action in superspace is 
\begin{equation} 
\label{SYM}
\mathcal{S}_{SYM}=\int {\rm d}^{4}x \mathcal{G}_{SYM}\left\{\frac{1}{32\pi}{\rm Im} 
\left [\tau \int {\rm d}^{2}\theta {\rm d}^{2}\bar{\theta} 
(\mathcal{K}_{{\rm NLYM}}(W^{1}_{\alpha},W^{2}_{\alpha})+{\rm h.c.})\right]\right\},
\end{equation}
where $\mathcal{G}_{NLSYM}\{\mathcal{O}\}$ and $\mathcal{K}_{{\rm NLYM}}$ are realizing the nonlocal deformation of the 
standard gauge-invariant term $W^{\alpha}W_{\alpha}$.
We can also consider a coupling to a matter chiral superfield as 
\begin{eqnarray} 
\label{SYMNLL}
\mathcal{S}=\mathcal{S}_{{\rm SYM}}+\int {\rm d}^{4}x\left(\mathcal{G}_{NLGT}
\left\{\int {\rm d}^{2}\theta {\rm d}^{2}\bar{\theta} 
\mathcal{K}_{{\rm NLM}}(\bar{\Phi}_{1}+\bar{\Phi}_{2},{\rm e}^{V_{1}}+{\rm e}^{V_{2}},
\Phi_{1}+\Phi_{2})\right\}\right)
\nonumber \\
+\int {\rm d}^{4}x \left[\int {\rm d}^{2}\theta \mathcal{W}(\Phi_{1}+\Phi_{2})
+\int {\rm d}^{2}\bar{\theta} \bar{\mathcal{W}}(\bar{\Phi}_{1}+\bar{\Phi}_{2})\right],
\end{eqnarray}
where $\mathcal{K}_{{\rm NLM}}$ and $\mathcal{G}_{NLM}\{\mathcal{O}\}$ are 
extending the standard term $\bar{\Phi}{\rm e}^{V}\Phi$ to a nonlocal one. 
A function $\mathcal{G}_{NLM}$ similar to the one considered above for Wess-Zumino
can be suggested in this case.

\subsection{Acausalities in super Yang-Mills}

In nonlocal supersymmetric Yang-Mills, in BRST quantization, the propagators are exactly the same 
as the ones just considered in the Wess-Zumino model, but with an extra color factor $\delta_{ab}$ 
(see, for nonsupersymmetric Yang-Mills, Ref. \cite{W3}). 
Note that we will have also ghost and auxiliary ghost propagators
\begin{equation} 
\label{pr1}
-\frac{{\rm i}\delta_{ab}}{(p^{2}-{\rm i}\epsilon)}{\rm exp}\left(-\frac{p^{2}}{\Lambda^{2}} \right)
=-{\rm i}\delta_{ab}\int_{1}^{\infty}\frac{{\rm d}\zeta}{\Lambda^{2}}{\rm exp}
\left(-\zeta \frac{p^{2}}{\Lambda^{2}} \right),
\end{equation}
\begin{equation} 
\label{pr1}
-\frac{{\rm i}\delta_{ab}}{(p^{2}-{\rm i}\epsilon)}\left\{1-{\rm exp}\left(-\frac{p^{2}}{\Lambda^{2}} 
\right)\right\}=-{\rm i}\delta_{ab}\int_{0}^{1}\frac{d\zeta}{\Lambda^{2}}
{\rm exp}\left(-\zeta \frac{p^{2}}{\Lambda^{2}} \right),
\end{equation}
and supersymmetric partners of ghosts, i.e., 
\begin{equation} 
\label{pr3}
-\frac{{\rm i}\delta_{\alpha \dot{\beta}}}{(\gamma_{\mu}p^{\mu}-{\rm i}\epsilon)}
{\rm exp}\left(-\frac{p^{2}}{\Lambda^{2}} \right),
\end{equation}
\begin{equation} 
\label{pr4}
-\frac{{\rm i}\delta_{\alpha \beta}}{(\gamma_{\mu}p^{\mu}-{\rm i}\epsilon)}
\left\{1-{\rm exp}\left(-\frac{p^{2}}{\Lambda^{2}} \right)\right\}.
\end{equation}
 
Let us consider, in analogy with the Wess-Zumino case, a scattering $A^{(1)}_{\mu}A^{(1)}_{\mu}
\rightarrow A^{(1)}_{\mu}A^{(1)}_{\mu}$. We have to note that now this interaction 
does not result from the $F$-terms as in Wess-Zumino $\phi^{(1)}\phi^{(1)}\rightarrow \phi^{(1)}\phi^{(1)}$, 
but from the $D$-terms. This holds despite 
the fact that the Feynman diagrams involved are exactly similar 
to Fig. 4 but with the replacements $\phi^{(1,2)} \rightarrow A_{\mu}^{(1,2)}$ 
and $\psi^{(1,2)}\rightarrow \lambda^{(1,2)}$
(and with extra color gauge factors in the vertices understood),
and with other extra contributions resulting from 
standard ghost supermultiplets and auxiliary ones. 

The nonrenormalization theorem does not protect the
$D$-terms from quantum loop corrections. 
As a consequence, supersymmetry is not the way to cancel acausalities in 
a generic super Yang-Mills theory. 
But generally, n-loops' diagrams resulting from the fermionic sector and the bosonic one
contribute to the divergences with opposite signs. 
Thus, we can mitigate the causality problem 
with a suitable gauge group and fields' content
in our theory. For example, with a tuning of the coupling parameters, 
in principle it could be possible to tune to zero the first relevant divergences 
in the total amplitudes. However, it is unclear how to achieve 
a cancellation to all orders in perturbation theory and 
to all orders of divergences  
with a finite content of fields. It therefore seems that the only way to 
obtain a complete cancellation of all acausal divergences 
in a nonlocal quantum field theory is to introduce 
an infinite number of fermions and bosons, but also in this framework 
it remains unclear whether exact cancellations can be proved.
On the other hand, we argue that also in this case
supersymmetry is protecting the model from acausal contributions 
in the $F$-terms. For this reason, supersymmetry seems 
a natural step towards the realization of a completely 
consistent nonlocal quantum field theory. 

\section{Conclusions and remarks}

The issue of nonlocality in field theory has many important aspects. In ordinary quantum field theory
on a flat spacetime background, the assumption of causality restricts the singularities of the $n$-point
Green functions to lie within the future tube. This means that the Green functions propagate positive
frequencies purely forward in time, and negative frequencies purely backwards. When spacetime is instead
curved, this picture may be substantially affected. For example, the Euclidean section of a black-hole
metric has nontrivial topology with Euler number $2$. This implies that there exists an obstruction to the 
Wick rotation of the time axis, and hence, when the Euclidean Green functions are analytically continued
to the Lorentzian regime, they contain acausal singularities periodically distributed in the imaginary
time coordinate \cite{Hawking1981}.
On the other hand, in a space-of-histories formulation of gauge theories, the connection forms 
naturally available are nonlocal, since they are defined with the help of Green functions of an
invertible operator acting on gauge fields, as is shown in detail in Ref. \cite{DeWitt}.

In this paper, we have shown how supersymmetry can play a fundamental role
for a consistent realization of nonlocal field theories.
In nonsupersymmetric $\lambda \phi^{4}$ models, acausal contributions occur 
at one loop in simple processes like $\phi\phi \rightarrow \phi\phi$ scatterings.
Also in the Wess-Zumino case, we have acausal divergent diagrams,
but the contributions resulting from the bosonic sector and the fermionic sector
cancel each other in $F$-terms, to all orders of perturbation theory. 
We remark that, in a nonlocal model, this corresponds 
to a complete cancellation of an infinite number of divergences.
We have also noticed that the nonrenormalization theorem of supersymmetry 
will guarantee us not only an alternative argument to the quantitative calculation 
given, but also a proof of cancellations to all orders. 
Unfortunately, the nonrenormalization theorem does not cancel 
all divergences: quantum corrections of $D$-terms are not protected 
and acasual divergences are not cancelled in corresponding diagrams. 
The results have been discussed also for a nonlocal supersymmetric Yang-Mills model, 
including the case of coupling to matter. Also in this last case, acausal contributions 
are cancelled in $F$-terms, but 
in $D$-terms they are generically remaining. 
The problem therefore arises to prove 
that, in order to cancel the infinitely many acausal diagrams, 
one has to introduce an infinite number of bosons and fermions.
This is naturally occurring in string theory, generically predicting a Kaluza-Klein tower
of modes, also higher spins' ones. 
It is quite natural to extend the formalism suggested in this paper
to the case of higher spins' superfields, in which generically 
the equations of motion have nonlocal kinetic terms, as for a spin-$3$ field. 
Thus, we think that our nonlocal supersymmetry model could be reinterpreted 
as an effective quantum field theory model of a string theory. 

On the other hand, supersymmetry does not occur at energies $E<{\rm TeV}$ and in principle,
if present, it could be broken near the Planck scale. Thus, nonlocal gauge theories continue to possess 
acausalities in quantum loops at energies lower than the supersymmetry-breaking scale. 
Of course, harmful loop diagrams resulting from $F$-terms are cancelled at $\Lambda_{{\rm SUSY}}$
when supersymmetry is restored; while the ones originating from $D$-terms 
are presumably alleviated at a scale lower than $\Lambda$ 
because of counter cancellations between bosons and fermions; 
and then completely cancelled at $\Lambda$, a process 
in which we think that an 
infinite number of bosons and fermions are excited. 
As argued in Refs. \cite{Joglekar1,Joglekar2,Joglekar3},
it is practically impossible to detect such violations 
if the $\Lambda$-scale of cutoff, introduced in the smeared propagators,
is (for example) close to the Planck scale. 

Another comment is regarding the problem of quantum gravity. 
In fact, as we said in the Introduction, practically all main candidates for a
quantum theory of gravity are predicting the loss of locality at the quantum scale.
This is a clear point in order to avoid an undesirable problem like the curvature singularity.
As a consequence, it is fundamental to consider a meaningful 
nonlocal quantum theory of matter, which makes it possible to develop perturbation theory. 
The present paper might represent a first step towards this goal: a cancellation of an infinite number of acausal 
infinities for a nonlocal quantum field theory. As far as the underlying classical theory
is concerned, it also remains to be seen how to formulate the Cauchy problem, a proper understanding 
of which is very important in theories of gravity or motivated by gravity \cite{Foures}. 

\acknowledgments
A. A. is grateful to the Galileo Galilei Institute for Theoretical physics 
for the hospitality, where this paper was prepared.
A. A. is particularly grateful to 
Massimo Bianchi for discussions and suggestions 
on these subjects. 
A.A. would also like to thank 
 Zurab Berezhiani, Francisco Morales, Luca Griguolo,
Gabriele Veneziano and  Anupam Mazumdar 
 for valuable remarks and suggestions.
G. E. is grateful to the Dipartimento di Fisica of Federico II University, Naples, for
hospitality and support, and to K. Kirsten for correspondence. The authors are indebted to
P. Townsend for a careful reading of the manuscript.

\begin{appendix}

\section{The nonrenormalization theorem of supersymmetry}

Since it plays a key role in our analysis, we now present a brief review of the conceptual framework for 
the nonrenormalization theorem of supersymmetry. A key idea \cite{NR2} is to think of all coupling
constants $\lambda_{i}$ in the superpotential as background chiral fields. The effective superpotential
of the dynamical fields $\phi_{I}$ and the background fields $\lambda_{i}$ is subject to the
following constraints \cite{NR2}:
\vskip 0.3cm
\noindent
(i) The nonvanishing values of coupling constants are interpreted as spontaneously breaking the
global symmetry group $G$. The effective Lagrangian which depends both on $\phi_{I}$ and 
$\lambda_{i}$ should be invariant under $G$.
\vskip 0.3cm
\noindent
(ii) The effective superpotential $W_{\rm eff}$ is a (locally) holomorphic function of all fields. 
Since the coupling constants are treated as fields as well, this implies that $W_{\rm eff}$ is
independent of the Hermitian conjugates $\lambda_{i}^{\dagger}$, unlike what happens in ordinary
field theories. 
\vskip 0.3cm
\noindent
(iii) The effective superpotential can depend on the dynamically generated scale of the theory,
denoted by $\Lambda$, and it should be smooth in the limit $\Lambda \rightarrow 0$. In almost all
cases, this implies that $W_{\rm eff}$ cannot grow faster than $\phi^{3}$, as a field $\phi$
takes increasingly large values. The gauge couplings are asymptotically free.
\vskip 0.3cm
\noindent
(iv) The behavior of $W_{\rm eff}$ when the coupling constants in the bare superpotential approach $0$
can be analyzed with perturbative methods, and this constrains the small $\lambda_{i}$ limit. It might
happen that there are more light fields at $\lambda_{i}=0$ than at nonvanishing values of $\lambda_{i}$.
Upon integrating out these fields and not including them in the effective action, $W_{\rm eff}$ might
be nonanalytic at $\lambda_{i}=0$. This is the peculiar weak-coupling regime. 

Interestingly, it turns out that the effective superpotential is not a generic function of the fields
consistent with the symmetries. There exist some terms which are consistent with all symmetries of the
problem but are not generated by perturbative or nonperturbative effects. This clearly violates the
so-called principle of naturalness. The author of Ref. \cite{NR2} develops heuristic arguments, and 
assumes that the theory can be regularized while preserving all symmetries.

We would like to note that nonrenormalization theorems of (rigid) supersymmetry 
are not lost upon relaxing the assumption of locality in a supersymmetric quantum field theory.
In fact, SuperPoincar\'e is a rigid global group of transformations,
as a generalization of the Poincar\'e one. A de-localization of an interaction spacetime point
preserves global supersymmetry. 
As a simple analogy, we consider scattering processes in classical mechanics:
in a collision between two, or N, pointlike masses,
angular momentum is conserved, as well as
in a collision between two, or N, rigid bodies;
{\it i.e.}, the angular momentum theorem
is not dependent on pointlike or volume-like interactions.
This is a general theorem of classical mechanics, as a consequence 
of Noether's theorem for translation and rotation global groups, 
in the Euclidean three-dimensional space. The Lagrangian of
classical mechanics is always invariant under the 
angular momentum operator $\vect{L}$, as well as the 
supersymmetry Lagrangian which is also invariant under the
$\vect{Q}$ generator. From this last trivial propriety it 
suddenly follows, in $\mathcal{N}=1$ supersymmetry, that only $D$- and $F$-terms can be written. 
Locality in the interactions is not relevant in this argument,  
in classical mechanics as well as in supersymmetric quantum field theory. 
On the other hand, for supergravity, local supersymmetry is requested, but a nonlocal supergravity 
is beyond the aims of this paper. 

\section{A brief review of a local Wess-Zumino model}

The Wess-Zumino model is a simple supersymmetric interacting theory
of a chiral supermultiplet $\Phi$ (see, for example, Ref. \cite{WB}). In the superspace formalism, we can write the 
Lagrangian of this model in an explicitly supersymmetric form \footnote{The kinetic term might be generalized to a 
more generic functional $\mathcal{K}(\Phi,\bar{\Phi})$ called nonminimal K\"{a}hler potential.} 
\begin{equation} 
\label{LSUSYWZ}
\mathcal{L}_{{\rm WZ}}=\mathcal{L}_{{\rm kin}}+\mathcal{L}_{{\rm int}},
\end{equation}
\begin{equation} 
\label{WZkin}
\mathcal{L}_{{\rm kin}}=\int {\rm d}^{2}\theta {\rm d}^{2}\bar{\theta} \bar{\Phi}\Phi+{\rm h.c.},
\end{equation}
\begin{equation} 
\label{WZint}
\mathcal{L}_{{\rm int}}=\int {\rm d}^{2}\theta \mathcal{W}(\Phi)
+\int {\rm d}^{2}\bar{\theta} \bar{\mathcal{W}}(\bar{\Phi})
=-\frac{\delta \mathcal{W}}{\delta \phi}F-\frac{1}{2}\frac{\delta^{2}\mathcal{W}}
{\delta \phi^{2}}\psi \psi +{\rm h.c.}, 
\end{equation}
where
\begin{equation} 
\label{Wphi}
\mathcal{W}(\Phi)=\mathcal{W}(\phi)+\sqrt{2}\frac{\delta \mathcal{W}}{\delta \phi}\theta \psi 
-\theta \theta \left(\frac{\delta \mathcal{W}}{\delta \phi}F
+\frac{1}{2}\frac{\delta^{2}\mathcal{W}}{\delta \phi^{2}}\psi \psi \right),
\end{equation}
having denoted by $F$ and $\bar{F}$ the auxiliary fields.
After integrating out the auxiliary fields, we obtain 
\begin{equation} 
\label{FFbar}
F=\frac{\delta \bar{\mathcal{W}}}{\delta \phi},\,\,\,\,\,
{\bar F}=\frac{\delta \mathcal{W}}{\delta \phi}.   
\end{equation}
Upon integrating over the Grassmannian superspace, the total Lagrangian is 
\begin{equation} 
\label{Lag}
\mathcal{L}_{\rm WZ}=\mathcal{L}_{{\rm kin}}+\mathcal{L}_{{\rm int}}
=\partial_{\mu} \bar{\phi}\partial^{\mu}\phi+\frac{{\rm i}}{2}\psi\partial_{\mu}\sigma^{\mu}\psi+\bar{F}F
-\frac{\delta \mathcal{W}}{\delta \phi}F
-\frac{1}{2}\frac{\delta^{2}W}{\delta \phi^{2}}\psi \psi+{\rm h.c.}
\end{equation}
For a renormalizable theory the superpotential is 
\begin{equation} 
\label{Wphii}
\mathcal{W}(\Phi^{i})=\frac{1}{2}m_{ij}\Phi^{i}\Phi^{j}+\frac{1}{3}g_{ijk}\Phi^{i}\Phi^{j}\Phi^{k}.
\end{equation}
We do not consider linear terms as usually done, but in principle this action can be extended 
with an extra term $a_{i}\Phi^{i}$. From Eqs. (\ref{Wphii}) and (\ref{Lag}), we obtain the Lagrangian terms
\begin{equation} 
\label{kin}
\mathcal{L}_{{\rm kin}}=\partial^{\mu}\phi^{*}\partial_{\mu}\phi
+\frac{{\rm i}}{2}\psi \gamma^{\mu}\partial_{\mu}\psi+{\rm h.c.},
\end{equation}
\begin{equation} 
\label{int}
\mathcal{L}_{{\rm int}}=-|m\phi +\lambda \phi^{2}|^{2}-\frac{1}{2}[m(\bar{\psi}_{R}\psi_{L} 
+ \bar{\psi}_{L}\psi_{R})+2\lambda(\phi \bar{\psi}_{R}\psi_{L}+\phi^{*}\bar{\psi}_{L}\psi_{R})].
\end{equation}

\section{Local super Yang-Mills model}

We here consider a supersymmetric Yang-Mills model, with locality, 
of a vector supermultiplet $V$ with a generic gauge group $SU(N)$ (see, for a general reference, 
Ref. \cite{WB}). The Lagrangian is
\begin{equation} 
\label{gauge}
\mathcal{L}_{SYM}=\frac{1}{32\pi}{\rm Im}\left( \tau\int d^{2}\theta W^{\alpha}W_{\alpha}\right)
={\rm Tr} \left[-\frac{1}{4}F_{\mu\nu}F^{\mu\nu}-i\lambda \sigma^{\mu}D_{\mu}\bar{\lambda}+\frac{1}{2}D^{2} \right]
+\frac{\theta_{YM}}{32\pi^{2}}g^{2}{\rm Tr} F_{\mu\nu}\tilde{F}^{\mu\nu},
\end{equation}
where 
\begin{equation} 
\label{Walpha}
W_{\alpha}=-\frac{1}{4}\bar{D}\bar{D}({\rm e}^{-V}D_{\alpha}{\rm e}^{V}),\,\,\,\,
\bar{W}_{\dot{\alpha}}=-\frac{1}{4}DD({\rm e}^{V}\bar{D}_{\alpha}{\rm e}^{-V}),
\end{equation}
having defined $\tau \equiv \frac{\theta_{YM}}{2\pi}+\frac{4\pi {\rm i}}{g^{2}}$ 
and $\tilde{F}^{\mu\nu} \equiv \frac{1}{2}\epsilon^{\mu\nu\rho\sigma}F_{\rho\sigma}$.

On considering a theory with matter chiral superfields coupled with the vector superfield,
the Lagrangian term (\ref{WZkin}) is extended as supersymmetric gauge-invariant term $\bar{\Phi}{\rm e}^{V}\Phi$, 
\begin{equation} 
\label{matter1}
\mathcal{L}_{{\rm matter}}=\int {\rm d}^{2}\bar{\theta}\bar{\Phi}{\rm e}^{V}\Phi
+\int {\rm d}^{2}\theta \mathcal{W}(\Phi)+\int {\rm d}^{2}\bar{\theta}\bar{W}(\bar{\Phi}).
\end{equation}
In the Wess-Zumino gauge 
\begin{equation} 
\label{eV}
\bar{\Phi}{\rm e}^{V}\Phi=\bar{\Phi}\Phi+\bar{\Phi}V\Phi+\frac{1}{2}\bar{\Phi}V^{2}\Phi ,
\end{equation}
\begin{equation} 
\label{V1}
[\bar{\Phi}V\Phi]_{\theta \theta \bar{\theta} \bar{\theta}}=\frac{{\rm i}}{2}\bar{\phi}A^{\mu}\partial_{\mu}\phi
-\frac{{\rm i}}{2}\partial_{\mu}\bar{\phi}A^{\mu}\phi
-\frac{1}{2}\bar{\psi}\bar{\sigma}^{\mu}A_{\mu}\psi+\frac{{\rm i}}{\sqrt{2}}\bar{\phi}\lambda \psi
-\frac{{\rm i}}{\sqrt{2}}\bar{\psi}\bar{\lambda}\phi+\frac{1}{2}\bar{\phi}D\phi ,
\end{equation}
\begin{equation} 
\label{V2}
[\bar{\Phi}V^{2}\Phi]_{\theta\theta \bar{\theta} \bar{\theta}}=\frac{1}{2}\bar{\phi}A^{\mu}A_{\mu}\phi ,
\end{equation}
and replacing $V\rightarrow 2gV$, we obtain a Lagrangian 
\begin{eqnarray} 
\label{SYM}
\mathcal{L}&=& \mathcal{L}_{{\rm SYM}}+\mathcal{L}_{{\rm matter}}+\mathcal{L}_{{\rm FI}}
=\frac{1}{32\pi}{\rm Im}\left(\tau \int {\rm d}^{2}\theta W^{\alpha}W_{\alpha} \right)
+2g\sum_{A}\zeta_{A}\int {\rm d}^{2}\theta {\rm d}^{2}\bar{\theta}V^{A}
\nonumber \\
&+& \int {\rm d}^{2}\theta {\rm d}^{2}\bar{\theta}\bar{\Phi}{\rm e}^{2gV}\Phi
+\int {\rm d}^{2}\theta \mathcal{W}(\Phi)+\int {\rm d}^{2}\bar{\theta}\bar{\mathcal{W}}(\bar{\Phi})
\nonumber \\
&=& {\rm Tr}\left[-\frac{1}{4}F_{\mu\nu}F^{\mu\nu}-{\rm i}\lambda \sigma^{\mu}D_{\mu}\bar{\lambda}
+\frac{1}{2}D^{2} \right]+\frac{\theta_{{\rm YM}}}{32\pi^{2}}g^{2}{\rm Tr} F_{\mu\nu}\tilde{F}^{\mu\nu}
\nonumber \\
&+& g\sum_{A}\zeta_{A}D^{A}+\bar{D}_{\mu}\bar{\phi}D^{\mu}\phi
-{\rm i} \psi \sigma^{\mu}D_{\mu}\bar{\psi}+\bar{F}F+{\rm i}\sqrt{2}g\bar{\phi}\lambda \psi
\nonumber \\
&-& {\rm i}\sqrt{2}g\bar{\psi}\bar{\lambda}\phi+g\bar{\phi}D\phi
-\frac{\delta \mathcal{W}}{\delta \phi^{i}}F^{i}
-\frac{\delta \bar{\mathcal{W}}}{\delta \bar{\phi}_{i}}\bar{F}_{i}
-\frac{1}{2}\frac{\delta^{2}\mathcal{W}}{\delta \phi^{i}\delta \phi^{j}}\psi^{i}\psi^{j}
-\frac{1}{2}\frac{\delta^{2}\bar{\mathcal{W}}}{\delta \bar{\phi}^{i}\delta \bar{\phi}^{j}}\bar{\psi}^{i}\bar{\psi}^{j},
\end{eqnarray}
where $D^{a}$ and $F^{i}$ are auxiliary fields satisfying the equations 
\begin{equation} 
\label{FWp}
D^{a}=-g\bar{\phi}T^{a}\phi-g\zeta^{a}, \;
{\bar F}_{i}=\frac{\delta \mathcal{W}}{\delta \phi^{i}},
\end{equation}
and $\mathcal{L}_{{\rm FI}}$ is the Fayet-Iliopoulos term.

As a final comment of this Appendix, 
the information about the group symmetry generators $T^{a}$ will be contained 
in the kinetic terms of the fields, or in the term $\bar{\Phi} {\rm e}^{2gV}\Phi$. 

\section{EWMKE's model: nonlocal $\lambda \phi^{4}$ theory}

In this Appendix, we briefly review the EWMKE ({\it i.e.}, Eliezer, Woodard, Moffat, Kleppe, Evens)   
model of a nonlocal and nonsupersymmetric $\lambda \phi^{4}$ theory. Within this framework,
the action functional can be split as
\begin{equation}
\label{W1}
\mathcal{S}[\phi]=\mathcal{F}[\phi]+\mathcal{I}[\phi],
\end{equation}
where $\mathcal{F}[\phi]$ is the free part, while $\mathcal{I}[\phi]$ is the interaction part.
$\mathcal{I}[\phi]$ is supposed to be analytic around the vacuum,
and $\mathcal{F}$ takes the form
\begin{equation}
\label{W2}
\mathcal{F}[\phi]=\frac{1}{2}\int {\rm d}^{D}x\phi_{i}F_{ij}\phi_{j}.
\end{equation}
The action $S$ acquires nonlocal nature through a smearing operator $\mathcal{E}$.
The EWMKE choice is 
\begin{equation}
\label{W3}
\mathcal{E}=\rm exp\left[\frac{F}{2\Lambda^{2}}\right],
\end{equation}
whit $\Lambda$ a cutoff scale of new physics beyond
effective quantum field theories. The fields $\phi$ are smeared according to 
\begin{equation}
\label{W2}
\hat{\phi}_{i}=\mathcal{E}_{ij}^{-1}\phi_{j}.
\end{equation}
We also define the operator 
\begin{equation}
\label{W3}
\mathcal{O} \equiv (\mathcal{E}^{2}-I)F^{-1}.
\end{equation}
At this stage, we introduce an auxiliary field $\varphi_{i}$ for each matter field $\phi_{i}$ in the form
\begin{equation}
\label{W4}
\mathcal{S}[\phi,\varphi]=\mathcal{F}[\hat{\phi}]-\mathcal{A}[\varphi]+\mathcal{I}[\phi+\varphi],
\end{equation}
\begin{equation}
\label{W5}
\mathcal{A}[\varphi]=\frac{1}{2}\int {\rm d}^{D}x\varphi_{i} \mathcal{O}_{ij}\varphi_{j}.
\end{equation}
The classical shadow field equation is
\begin{equation}
\label{W6}
\frac{\delta \mathcal{S}[\phi, \varphi]}{\delta \varphi(x)}=0.
\end{equation}
The nonlocal action is obtained by substituting the solution of the 
classical field equation, Eq. (\ref{W6}), into Eq. (\ref{W4}), {\it i.e.}, by substituting 
$\varphi=\mathcal{O}_{ij} {\delta \mathcal{I}[\phi+\varphi] \over \delta \varphi_{j}}$. 

\section{Quantization in the EWMKE model}

In this section, we briefly review quantization issues,
as discussed in Refs. \cite{W1,W2,Moffat,Evens,W3}.
Consider therefore the vacuum expectation value of an arbitrary operator 
$\mathcal{O}$ as
\begin{equation}
\label{quantization}
<\mathcal{T}\{\mathcal{O}[\phi]\}>_{\mathcal{E}}=\int \mathcal{D}\phi\, \mu[\phi]
\left(G.F.\right)\,\mathcal{O}[\hat{\phi}]\,\rm exp\{i\hat{S}[\phi]\},
\end{equation}
where $\mathcal{T}$ is the time-ordering operator, and $G.F.$ is the gauge fixing. 
$\mathcal{O}$ is nonlocally regulated in a nonlocal theory.
Eq. (\ref{quantization}) defines the quantization.
The problem of a consistent quantization of a nonlocal field theory
is reduced to the problem of existence of 
the measure factor $\mu[\phi]$ and the gauge fixing.
These are necessary in order to preserve unitarity. 

The perturbative unitarity of a nonlocal quantum field theory 
was discussed in papers cited above, applying the Cutkosky rules
to EWMKE nonlocal theories where poles are only the zeros of 
$F=0$, and where interaction vertices are functions of $F$ only.
Under these quite general requirements, unitarity on 
a large subspace of states $\mathcal{M}$, in the Fock space,
was shown. $\mathcal{M}$ can also contain ghost fields of BRST
quantization, as unphysical polarizations. 
Gauge invariance guarantees the ghosts' decoupling  
and unitarity. In the nonlocal procedure, 
we start from a local quantum field theory, smearing fields
and delocalizing vertices. If a continuos transformation
$\delta \phi^{1}_{i}=T_{i}[\phi^{1}]$
generates a symmetry of the local action $S[\phi^{1}]$, 
the corresponding transformation of the nonlocal one is  
$\hat{\delta}\phi^{1}_{i}=\mathcal{E}_{ij}^{2}T_{j}[\phi^{1}+\phi^{2}[\phi^{1}]]$, as mentioned above. 
Such a symmetry has to preserve $\mathcal{D}\phi\mu[\phi]$ of the
nonlocal quantum field theory. This corresponds to the following relation:
\begin{equation} \label{conq}
\hat{\delta}\left[ {\rm log}(\mu[\phi]) \right]=-{\rm Tr}\left[\frac{\delta \hat{\delta} \phi_{i}}
{\delta \phi_{j}}\right]=-{\rm Tr} \left[\mathcal{E}_{ik}^{2}\frac{\delta T_{k}}
{\delta \phi_{l}}[\phi_{1}+\phi_{2}[\phi_{1}]]K_{lk}[\phi_{1}+\phi_{2}[\phi_{1}]]\mathcal{O}_{kj}^{-1} \right].
\end{equation}

Under this quantization procedure, we can recover 
Feynman rules of $\hat{S}[\phi_{1},\phi_{2}]$ as simple extension of usual ones:
propagators are smeared by a factor $\mathcal{E}^{2}$, and vertices remains the same,
as mentioned above. The $\phi_{2}$ are auxiliary fields 
propagating only off-shell, because they are projected-out 
by solutions of classical field equations $\phi_{2}[\phi_{1}]$. 
All details are extensively discussed in papers 
just cited above in this same appendix. 

\end{appendix}

\end{document}